\documentclass[10pt]{article}
\usepackage{fullpage,citesort,epsfig,graphics,amsbsy}
\usepackage{psfrag}
\usepackage[normal,small]{caption}
\newcommand{\beq}{\begin{equation}}
\newcommand{\eeq}{\end{equation}}
\newcommand{\bea}{\begin{eqnarray}}
\newcommand{\eea}{\end{eqnarray}}

\newcommand{\be}{\begin{equation}}
\newcommand{\ee}{\end{equation}}
\newcommand{\bq}{\begin{eqnarray}}
\newcommand{\eq}{\end{eqnarray}}
\newcommand{\ket}[1]{|#1\rangle}

\def\math{\mathsurround=0pt }
\def\leftrightarrowfill{$\math \mathord\leftarrow \mkern-6mu \cleaders\hbox{$\mkern-2mu \mathord- \mkern-2mu$}\hfill
 \mkern-6mu \mathord\rightarrow$}
\def\overleftrightarrow#1{\vbox{\ialign{##\crcr
     \leftrightarrowfill\crcr\noalign{\kern-1pt\nointerlineskip}
     $\hfil\displaystyle{#1}\hfil$\crcr}}}

\newcommand{\bfs}{\boldsymbol}

\let\l=\lambda    \let\p=\pi

 \def\bd{\begin{document}} \def\ed{\end{document}}
\def\ds{\documentstyle} \let\fr=\frac \let\bl=\bigl \let\br=\bigr
\let\Br=\Bigr \let\Bl=\Bigl
\let\bm=\bibitem
\let\na=\nabla
\let\pa=\partial \let\ov=\overline
\def\ft#1#2{{\textstyle{{\scriptstyle #1}\over {\scriptstyle #2}}}}
\def\fft#1#2{{#1 \over #2}}
\def\vp{\varphi}
\def\sst#1{{\scriptscriptstyle #1}}
\def\oneone{\rlap 1\mkern4mu{\rm l}}
\def\td{\tilde}
\def\wtd{\widetilde}
\def\dalemb#1#2{{\vbox{\hrule height .#2pt
        \hbox{\vrule width.#2pt height#1pt \kern#1pt
                \vrule width.#2pt}
        \hrule height.#2pt}}}
\def\square{\mathord{\dalemb{6.8}{7}\hbox{\hskip1pt}}}
\def\wtd{\widetilde}
\def\R{\rlap{\rm I}\mkern3mu{\rm R}}
\def\im{{\rm i}}
\def\tilg{\tilde{g}}
\def\tilF{\tilde{F}}
\def\tilA{\tilde{A}}
\def\varf{\varphi}
\def\tilf{\tilde{\phi}}
\def\tilh{\tilde{h}}
\def\rme{{\rm e}}
\def\ep{\epsilon}
\def\0{{(0)}}
\def\9{{(9)}}
\def\8{{(8)}}
\def\7{{(7)}}
\def\6{{(6)}}
\def\5{{(5)}}
\def\4{{(4)}}
\def\3{{(3)}}
\def\2{{(2)}}
\def\1{{(1)}}
\newcommand{\trace}{{\rm Tr}}
\newcommand{\ub}{\overline{U}}
\newcommand{\vb}{\overline{V}}
\newcommand{\uh}{\widehat{U}}
\newcommand{\vh}{\widehat{V}}
\newcommand{\ubh}{\overline{\widehat{U}}}
\newcommand{\vbh}{\overline{\widehat{V}}}
\newcommand{\lb}{\bar{\l}}
\newcommand{\Fb}{\overline{F}}
\newcommand{\Fh}{\widehat{F}}
\newcommand{\Fbh}{\overline{\widehat{F}}}
\newcommand{\Ab}{\overline{A}}
\newcommand{\Ah}{\widehat{A}}
\newcommand{\Abh}{\overline{\widehat{A}}}
\newcommand{\Gb}{\overline{G}}
\newcommand{\Gh}{\widehat{G}}
\newcommand{\Gbh}{\overline{\widehat{G}}}
\newcommand{\Pb}{\overline{P}}
\newcommand{\Ph}{\widehat{P}}
\newcommand{\Pbh}{\overline{\widehat{P}}}
\newcommand{\Qb}{\overline{Q}}
\newcommand{\Qh}{\widehat{Q}}
\newcommand{\Qbh}{\overline{\widehat{Q}}}
\newcommand{\Bb}{\overline{B}}
\newcommand{\Bh}{\widehat{B}}
\newcommand{\Bbh}{\overline{\widehat{B}}}
\newcommand{\fhns}{\hat{F}^{\rm (NS)}}
\newcommand{\fhrr}{\hat{F}^{\rm (RR)}}
\newcommand{\ahns}{\hat{A}^{\rm (NS)}}
\newcommand{\ahrr}{\hat{A}^{\rm (RR)}}
\newcommand{\hhrr}{\hat{H}^{\rm (RR)}}
\newcommand{\hchi}{\hat{\chi}}
\newcommand{\hphi}{\hat{\phi}}
\newcommand{\htau}{\hat{\tau}}
\newcommand{\cG}{{\cal G}}
\newcommand{\cGb}{\overline{{\cal G}}}
\newcommand{\cH}{{\cal H}}
\newcommand{\cP}{{\cal P}}
\newcommand{\cPb}{\overline{{\cal P}}}
\newcommand{\cQ}{{\cal Q}}
\newcommand{\cQb}{\overline{{\cal Q}}}
\newcommand{\cM}{{\cal M}}
\newcommand{\cN}{{\cal N}}
\newcommand{\cO}{{\cal O}}
\newcommand{\cD}{{\cal D}}
\newcommand{\cL}{{\cal L}}
\newcommand{\cA}{{\cal A}}
\newcommand{\cB}{{\cal B}}
\newcommand{\hg}{\hat{g}}
\newcommand{\cE}{{\cal E}}

\newcommand{\vpp}{\mbox{$\langle{\scriptstyle++}\rangle$}}
\newcommand{\vmp}{\mbox{$\langle{\scriptstyle-+}\rangle$}}
\newcommand{\vppp}{\mbox{$\langle{\scriptstyle+++}\rangle$}}
\newcommand{\vmpp}{\mbox{$\langle{\scriptstyle-++}\rangle$}}
\newcommand{\vpmp}{\mbox{$\langle{\scriptstyle+-+}\rangle$}}

\begin{document}
\setlength{\captionmargin}{20pt}
\begin{titlepage}
\begin{flushright}
UFIFT-HEP-05-17
\end{flushright}

\vskip 3cm

\begin{center}
\begin{Large}
{\bf Scattering of Glue by Glue on the Light-cone Worldsheet I:
Helicity  Non-conserving Amplitudes \footnote{Supported 
in part by the Department
of Energy under Grant No. DE-FG02-97ER-41029.}}
\end{Large}

\vskip 2cm
{\large 
D. Chakrabarti\footnote{E-mail  address: {\tt dipankar@phys.ufl.edu}}, 
J. Qiu\footnote{E-mail  address: {\tt jqiu@phys.ufl.edu}}, 
and C. B. Thorn\footnote{E-mail  address: {\tt thorn@phys.ufl.edu}}
}
\vskip0.20cm
{\it Institute for Fundamental Theory\\
Department of Physics, University of Florida,
Gainesville FL 32611}


\vskip 1.0cm
\end{center}

\begin{abstract}\noindent
We give the light-cone gauge calculation of the one-loop on-shell scattering
amplitudes for gluon-gluon scattering
which violate helicity conservation. 
We regulate infrared divergences by discretizing the $p^+$
integrations, omitting the terms with $p^+=0$. Collinear divergences
are absent diagram by diagram for the helicity non-conserving
amplitudes. We also employ
a novel ultraviolet regulator that is
natural for the light-cone worldsheet description
of planar Feynman diagrams. We show that
these regulators give the known answers for the helicity
non-conserving one-loop amplitudes, which
don't suffer from the usual infrared
vagaries of massless particle scattering. 
For the maximal helicity violating process we elucidate 
the physics of the remarkable fact that the loop momentum
integrand for the on-shell Green function associated
with this process,  
with a suitable momentum routing of the different 
contributing topologies, is identically zero. 
We enumerate the counterterms that must be included to give Lorentz 
covariant results to this order, and we show that they can
be described locally in the light-cone worldsheet formulation of
the sum of planar diagrams.
\end{abstract}
\vfill
\end{titlepage}
\section{Introduction and Conclusion}
The lightcone worldsheet representation of the sum of
the planar diagrams of quantum field theory \cite{thooftlargen,bardakcit}
shows a sense in which the string/field duality proposed by
Maldacena \cite{maldacena} for the case of certain supersymmetric gauge
theories can be universally valid for almost any field theory.
However, unlike the ${\cal N}=4$ supersymmetric gauge theory initially
studied by Maldacena, a general quantum field theory has ultraviolet
divergences that can complicate (or possibly ruin) the string description.
Indeed the generic worldsheet representation proposed in
$\cite{bardakcit}$ applies directly only to the bare diagrams
of the quantum field theory, and one must apply an explicit $UV$ cutoff
to give it concrete meaning. Since any physical cutoff will break
Lorentz invariance, there is the danger that the counterterms
required to restore it may not be compatible with
a {\it local} worldsheet description.

For the case of cubic scalar field theory in six space-time
dimensions (the marginally renormalizable situation), 
a study of the divergence structure has confirmed that
every counterterm required for the restoration of Lorentz invariance
has a viable local worldsheet
interpretation \cite{thornscalar}. In that work, it was shown that
two counterterms beyond the ones associated with mass, wave-function,
and coupling renormalization were required. But it turned out that
they only contributed to the self-energy diagrams: one could be
interpreted as a simple rescaling of the worldsheet action (or
alternatively as a renormalization of the speed of light), while the
other new counterterm represented a modification of the 
boundary conditions enjoyed by the worldsheet fields. Both of these
new counterterms are compatible with a local worldsheet dynamics.

It would be highly desirable to extend this all orders
conclusion to the worldsheet description of
gauge theories, such as the gauge sector of QCD
\cite{thornsheet}.
Unfortunately, because we work in light-cone gauge, 
the corresponding analysis is
considerably more complex, and we are still short of a complete
all orders result. However, we have completed the one loop analysis
and describe it in a series of two articles.
In this one we calculate the helicity amplitudes with
helicity nonconservation. Since the tree amplitudes for
such processes vanish, the one-loop amplitudes are finite in 
both the ultraviolet and infrared. In the sequel we shall deal with
the helicity conserving amplitudes which suffer from infrared
divergences.

While the literature has 
dealt with one-loop coupling renormalization in light-cone gauge 
including confirmation of asymptotic freedom \cite{thornfreedom,perryfreedom},
a check of Lorentz invariance at one loop requires the complete evaluation,
including all finite contributions, of the amplitudes for
a manifestly physical scattering process. The simplest
such process is the on-shell scattering of glue
by glue, and a complete light-cone gauge calculation of
this process is, to our knowledge, unavailable. We offer such a
calculation in this work, using the worldsheet friendly ultraviolet
cutoff employed in \cite{thornscalar}. 

\begin{figure}[ht]
\psfrag{'k1'}{$k_1$}
\psfrag{'k2'}{$k_2$}
\psfrag{'k3'}{$k_3$}
\psfrag{'k0'}{$k_0$}
\psfrag{'q1'}{$q_1$}
\psfrag{'q2'}{$q_2$}
\psfrag{'q3'}{$q_3$}
\psfrag{'q4'}{$q_4$}
\begin{center}
\includegraphics[width=7cm]{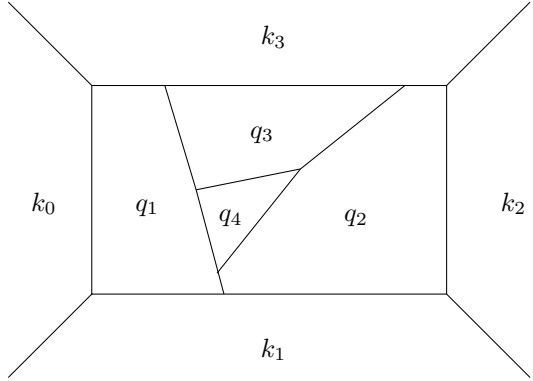}
\caption{Assignment of dual momentum variables to a planar diagram.}
\label{dualmom}
\end{center}
\end{figure}

For the reader unfamiliar with \cite{thornscalar} we briefly
describe here how this cutoff is implemented. The worldsheet
formalism maps the sum of planar diagrams (singled
out in $SU(N_c)$ gauge theory, for example, by 't Hooft's large $N_c$ limit 
\cite{thooftlargen}) to a rectangular
worldsheet whose length and width are light-cone time 
and longitudinal momentum respectively. 
To study this mapping in perturbation theory, it suffices to
use a regulator that exploits the planarity of the diagrams.
A general planar diagram can be drawn on a plane with
external lines extending to infinity (see Fig.~\ref{dualmom}). 
The internal and external
lines divide the plane into regions. Loops bound finite
regions in the plane and external lines bound regions extending to
infinity. The worldsheet description of planar diagrams 
\cite{thooftlargen,bardakcit}
employs ``dual momentum'' variables, $q_i$ for finite regions
and $k_j$ for external regions, one for each region as shown
in Fig.~\ref{dualmom}. 
The actual momentum carried by any line, which separates
two regions of the plane, is the difference
of the dual momentum variables of those regions\footnote{Since
$SU(N_c)$ planar diagrams are orientable, we can establish
a unique convention that the momentum carried by the line in a given direction
is the dual momentum of the region
to the right minus that of the region to the left with respect to
someone facing in the direction of momentum flow.}. The Feynman diagram
is given by an integral over all the $q_i$. In these
variables, the worldsheet friendly ultraviolet cutoff $\delta>0$
described in \cite{thornscalar} is implemented by simply inserting a
factor $e^{-\delta\sum_i{\bfs q}^2_i}$ in the loop integrand, 
where ${\bfs q}_i$ is the transverse dual momentum of the 
region bounded by loop $i$. Note that it only cuts off transverse
momentum. This is because the worldsheet formalism promotes
the transverse dual momenta to worldsheet fields whereas the $\pm$
coordinates merely parametrize the worldsheet.

As already mentioned the physical quantities we calculate
in this article are on shell scattering amplitudes.
In quantum field theories 
with massless fields (in particular gauge theories), these quantities
have infrared as well as ultraviolet divergences,
and we must also regulate these. It is quite natural to
define the worldsheet path integral on a lattice \cite{gilest,bardakcit},
and in the lightcone worldsheet description of field theories
such a lattice corresponds to discretizing
the $+$ component of momenta $p^+=lm$, $l=1,2,\ldots$,
where $p^\pm\equiv(p^0\pm p^z)/\sqrt2$
\cite{gilest,casher,thornfishnets}. We choose to do this, and thereby
we regulate all infrared divergences that come from integration 
regions where all components of a loop momentum are small. 
But scattering amplitudes also typically have collinear divergences from the
regions where an internal momentum is parallel to (collinear with) 
an on-shell external momentum. 
These divergences are not regulated by discretization of $p^+$.
This would be problematic for scalar couplings
but for gauge theories, in light-cone gauge, 
the cubic vertex provides an extra zero that renders the
problem manageable. Actually the easiest way to see this
uses the light-cone description itself. Consider an on-shell
external line with momentum $k$ hooked to two internal lines with
momenta $p,p-k$. 
In a frame where the transverse components ${\bfs k}=0$,
the propagators of the internal lines supply factors
\bea
{1\over p^2(p-k)^2}&=&{1\over ({\bfs p}^2-2p^+p^-)({\bfs p}^2
-2(p^+-k^+)p^-)}\nonumber
\eea
Integrating this expression near ${\bfs p}, p^-=0$ shows a logarithmic
divergence even if $p^+\neq 0, k^+$. Happily, the cubic gauge coupling
in lightcone gauge supplies a further factor
\bea
{\bfs K}=k^+{\bfs p}-p^+{\bfs k}\to k^+{\bfs p}\nonumber
\eea
which gives an additional zero that removes the divergence. As a
result the only impact that the collinear divergences have is
in the on-shell limit of self-energy corrections. The off-shell 
($k^2\neq0$) self-energy
correction has the small momentum behavior $O(k^2\ln k^2)$, where the 
logarithmic cut is associated with threshold for massless
gluon production, which occurs in the collinear limit. 
Since we divide by $k^2$ to get the
wave function renormalization, we see that the latter
will have a logarithmic collinear divergence in the on-shell limit
$k^2\to0$.
Thus, in lightcone gauge, the collinear divergence
problem is limited to self-energy bubbles on external lines.
But, for the helicity nonconserving amplitudes calculated in this
article, such diagrams don't contribute and the problem disappears.
In the second article of this series these contributions will
be present and will be tied up with the resolution of the
physics of soft gluon emission.

Before summarizing the content of this article,
there is a final general point about lightcone gauge we wish to mention
here. With this gauge choice, only the transverse components of
the gauge field ${\bfs A}$ remain as propagating degrees of
freedom. In 4 space-time dimensions,
it is convenient to assemble the two transverse components into
a complex field $A^\wedge\equiv(A^1+iA^2)/\sqrt2$, with complex conjugate  
$A^\vee\equiv(A^1-iA^2)/\sqrt2$ represented in Feynman diagrams by attaching an
arrow, representing a ``charge flow'' to each line of the diagram.
But what is the physical interpretation of this ``charge''? At first
glance it seems to be just the ``spin'' projected on the $z$-axis.
But this interpretation makes it seem highly frame dependent.
A not widely recognized fact is that for an on-shell gluon the ``charge'' 
is actually the frame independent helicity of the gluon.
The helicity $h={\vec p}\cdot{\vec J}/|{\vec p}|$ of a massless
particle is well-known to be a Lorentz invariant. (Here ${\vec p}$
is the three momentum of the gluon and ${\vec J}$ is the total 
angular momentum.). It is straightforward to express 
${\vec p}$ and ${\vec J}$
in terms of the light-cone components of the Poincar\'e generators
and apply $h$ to a single gluon state and find
\bea
h[a_1^\dagger({\bfs p},p^+)\pm ia_2^\dagger({\bfs p},p^+)]\ket{0}
=\pm[a_1^\dagger({\bfs p},p^+)\pm ia_2^\dagger({\bfs p},p^+)]\ket{0}
\nonumber\eea
It follows that the globally defined combinations
$(a^\dagger_1(p)\pm ia^\dagger_2(p))/\sqrt{2}$ create one
gluon states of helicity $\pm1$ no matter which direction the
gluon is moving! Thus the light-cone Feynman rules
in this complex field basis directly give the helicity scattering amplitudes
in the on-shell limit. 
This probably indicates a close connection of the light-cone worldsheet
\cite{bardakcit} description of gauge theory \cite{thornsheet} to
the twistor string representation \cite{wittentwistor,cachazosw} 
of gauge theory. In the latter work the helicity amplitudes
also play a central role. We think it likely that the lightcone
worldsheet formalism provides a concrete all orders lightcone
gauge fixed realization of the twistor string idea.

We now turn to a synopsis of the rest of the article. 
In Section 2 we quote the
Feynman rules in lightcone gauge. This gauge completely removes the redundant
gauge fields $A_\pm$ from the formalism, so the lines of every
diagram represent only the transverse components of the gluons. 
Central to the Feynman rules are the quantities $K^\mu_{ij}
=p_i^+p_j^\mu-p_j^+p_i^\mu$. In Section 3, we obtain a number of
identities enjoyed by the the $K_{ij}$ that enable dramatic
simplification of the final results for each helicity amplitude.
The quantities $K_{ij}$ play the role of the bispinor matrix
elements that occur in the famous Parke-Taylor formulas \cite{parketaylor}
for gluon tree amplitudes. 

Section 4 is devoted to the calculation of four gluon tree amplitudes.
Tree diagrams with four like helicities cannot even be drawn in
light-cone gauge, so their vanishing is guaranteed from the start.
The diagrams for tree amplitudes with three like-helicities {\it can} be
drawn and are non-zero off-shell. However the $K$ identities can be used to
show that they vanish on-shell. We also obtain the Parke-Taylor
forms for the helicity conserving tree amplitudes, though their one-loop
corrections are reserved for the second paper in this series.

Section 5 discusses the gluon self energy diagrams. These have been 
calculated previously in various treatments of lightcone gauge
but not with the worldsheet friendly regulator we are employing here.
Thus we give the calculation in complete detail. In particular we determine the
counterterms required to maintain Lorentz invariance and show that
they have a local worldsheet description.
In Section 6 we quote the results for the one-loop cubic vertex function
with calculational details to be found in \cite{thornlcnotes}.
Again we note the counterterms required by Lorentz invariance and
give their representation in the worldsheet formalism.

We discuss box diagrams in Section 7. For the helicity nonconserving
amplitudes discussed in this paper it turns out that the boxes
can all be reduced to a sum of triangle-like diagrams. We show
how this reduction takes place and quote the set of triangle-like
diagrams descended from the boxes in each case.

Finally in Sections 8 and 9 we put everything together and
calculate the on-shell amplitudes for the
scattering of glue by glue in the case of four like helicities and three
like helicities respectively. The final answer for
each of these cases is finite in both the ultraviolet and infrared and
agrees with the known answer.
In the four like helicity case, every one-loop diagram is
finite. However the
individual light-cone gauge diagrams combine in a very interesting way.
Bern \cite{bernprivate} has observed that, if one takes all contributing
diagrams to the on-shell one-loop four point Green function, and
routes momenta appropriately, the integrands for all these diagrams
sum to zero {\it before} integration. The appropriate momentum routing
turns out to be precisely that dictated by the worldsheet description
of planar diagrams \cite{bardakcit}. In spite of this
vanishing integrand, the physical scattering is non-zero, because the
helicity flipping self-energy insertions essential 
for this cancellation to occur are non-zero,
and must be removed by counterterms. Thus the physical scattering amplitude
is given by the negative of the sum of helicity flipping self energy 
insertions--a relatively simple calculation! 
This case serves as a nontrivial 
test of the
validity of one of the counterterms, namely the one that cancels
the helicity flipping component of the gluon self-energy diagram.
In contrast, the three like helicity case tests much more. The individual
diagrams contributing to this process are divergent in both the
ultraviolet and infrared. These divergences cancel in the
sum of all contributing diagrams, but all of the counterterms we have
determined {\it must be included to get the long known correct
answer} \cite{bernk,kunsztst}. We mention that our methods for obtaining
one-loop scattering amplitudes have
some similarity and overlap with those of \cite{berndk}. 

It is an important constraint on the all orders validity
of the worldsheet description of gauge theories \cite{bardakcit,thornsheet}
that {\it all} of the counterterms required to achieve Lorentz invariance
can be interpreted as new terms in a {\it local} worldsheet action.
Confirmation of this requirement to one-loop
order is the main achievement of the work described here. The second article in
this series will extend the one-loop analysis to the IR divergent
helicity conserving case. But a complete all orders demonstration
that all counterterms have a local lightcone worldsheet
interpretation  remains a distant goal.

\section{Feynman Rules for Light-cone gauge Yang-Mills}
We use the notation and conventions in Ref.~\cite{beringrt},
according to which the values of the 
non-vanishing three transverse gluon vertices are:
\bea
{{}\atop\mbox{\includegraphics[width=1.2cm]
{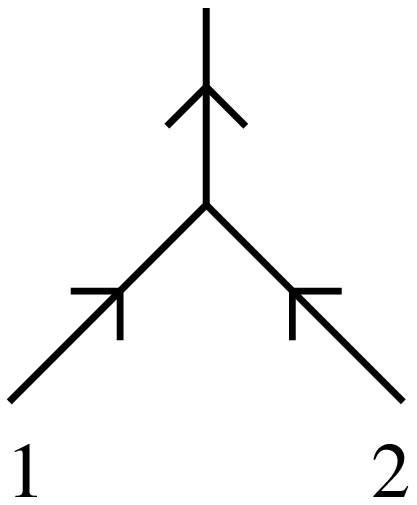}}}
\displaystyle\quad&=&{2gp_3^+\over p_1^+p_2^+}
\left(p_1^+{p_2^\wedge}-p_2^+{p_1^\wedge}\right)
={2gp_3^+\over p_1^+p_2^+}K^\wedge_{12}
\label{upupdown}\\
{{}\atop\mbox{\includegraphics[width=1.2cm]
{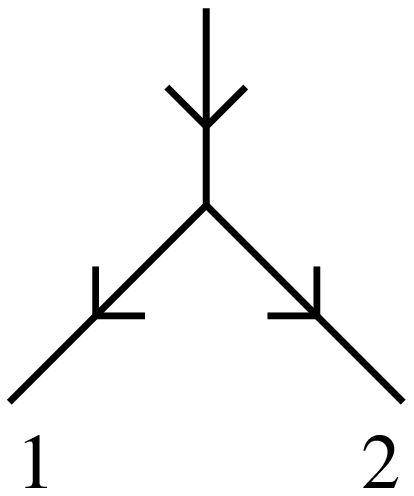}}}
\displaystyle\quad&=&{2gp_3^+\over p_1^+p_2^+}
\left(p_1^+{p_2^\vee}-p_2^+{p_1^\vee}\right)
={2gp_3^+\over p_1^+p_2^+}K^\vee_{12}
\label{downdownup}
\eea 
Here, $p_j^\wedge=(p_j^x+ip_j^y)/\sqrt2$, $p_j^\vee=(p_j^x-ip_j^y)/\sqrt2$, 
and $p_j^+$  are the components of the momentum 
{\it entering} the diagram on leg $j$. 
This coupling constant $g$ differs by a factor of $\sqrt{2}$ from the
conventional one $g=g_s/\sqrt2$.
We remind the reader that these are light-cone gauge ($A_-=0$) expressions
and include the contributions that arise when the longitudinal
field $A_+$ is eliminated from the formalism.

The quartic vertices in this helicity basis are given by
\bea
{{}\atop\mbox{\includegraphics[width=1.2cm]
{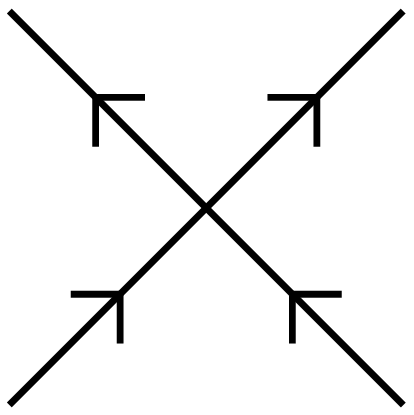}}}
\displaystyle\quad&=&-2{{g^2}{p^+_1p^+_3+p^+_2p^+_4\over
(p^+_1+p^+_4)^2}
\label{upupdowndown}}\\
{{}\atop\mbox{\includegraphics[width=1.2cm]
{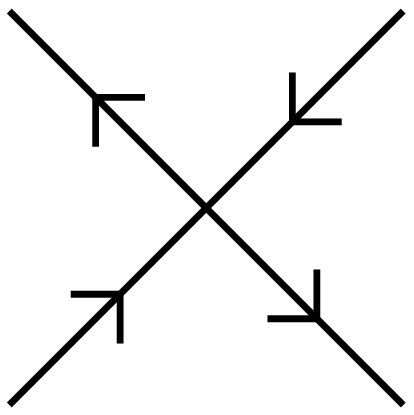}}}
\displaystyle\quad&=&+2{{g^2}\left({p^+_1p^+_2+p^+_3p_4^+\over
(p^+_1+p^+_4)^2}+{p^+_1p^+_4+p_2^+p^+_3\over
(p^+_1+p^+_2)^2}\right)}
\label{updownupdown}
\eea
where we again stress that these expressions 
include contributions from the elimination of $A_+$.
We also should point out that we are giving these rules in the context of the 
$1/N_c$ expansion, so the planar diagrams of the $SU(N_c)$ theory
are correctly given
with the simple substitution $g\to g\sqrt{N_c}$. Non-planar diagrams
after this substitution must be accompanied by appropriate
powers of $1/N^2_c$, depending on the
number of ``handles'' in the diagram. We have not spelled
that out here, because our focus will be on the planar diagrams
in this article. The results we obtain should therefore be compared
to the 't Hooft limit $N_c\to\infty$, fixed $g^2N_c$ of those in
the literature. In making such comparisons, note that the
substitution rule $g\to g\sqrt{N_c}$ multiplies conventionally
defined $n$-gluon tree amplitudes by a factor $N_c^{n/2-1}\to N_c$ for
$n=4$, so
for each gluon scattering process we remove this factor
before comparing to the literature.

\section{K Identities}
As we have seen the quantities
\bea
K^\mu_{ij}&\equiv& p^+_ip^\mu_j-p^+_jp^\mu_i
\eea
play a central role in the cubic Yang-Mills vertex. In fact,
we shall find that the simplest forms of the various helicity
amplitudes are achieved by expressing them as functions of the $K$'s.
These simple forms are in fact identical to those discovered by
Parke and Taylor using a bispinor
representation of polarization vectors \cite{parketaylor}. 
For us the role of the spinor matrix 
elements in those formulas will be played exclusively by $K_{ij}^\wedge$
and $K_{ij}^\vee$.

In order to reduce the expressions for the helicity amplitudes to the 
Parke-Taylor form, we will need a number of identities enjoyed by the $K$'s.
For a general $n$-gluon amplitude we can form $K_{ij}$ for each pair
of gluons $(ij)$, where $i,j=1,\ldots,n$ distinguish the different gluons.
By momentum conservation, it is immediate that
\bea
\sum_j K^\mu_{ij}&=&0.
\eea
From the fact that $K$ is an anti-symmetric product we have Bianchi-like
identities
\bea
p^+_iK^\mu_{jk}+p^+_kK^\mu_{ij}+p^+_jK^\mu_{ki}&=&0\\
K^\wedge_{li}K^\wedge_{jk}+K^\wedge_{lk}K^\wedge_{ij}
+K^\wedge_{lj}K^\wedge_{ki}&=&0
\eea
Finally, the most powerful type of identity follows from
a very simple calculation
\bea
\sum_j{K^\wedge_{ij}K^\vee_{jk}\over p^+_j}&=&-p^+_ip^+_k\sum_j 
{p_j^2\over2p^+_j}
\eea
which seems like a complicated non-linear relation. However 
when we are considering scattering amplitudes, the momenta 
all satisfy $p_i^2=0$ so the right side is zero! This identity plays a
central role in showing that trees with all
but one like helicities vanish. The $K$ identities
are also crucial for
reducing the complexity of the helicity amplitudes that don't vanish.

\section{Tree Amplitudes}

\begin{figure}[ht]
\begin{center}
\psfrag{'1'}{$1$}
\psfrag{'2'}{$2$}
\psfrag{'3'}{$3$}
\psfrag{'4'}{$4$}
\includegraphics[width=3.5in]{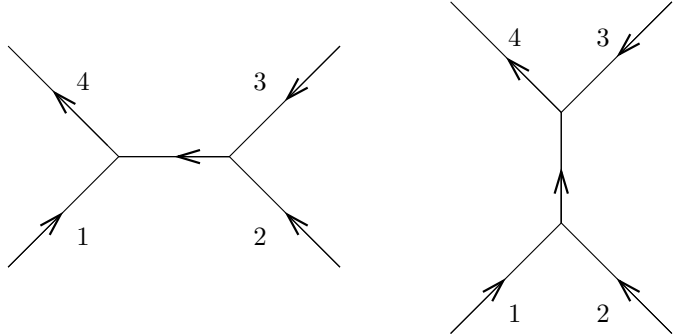}
\caption{Tree diagrams for the gluon scattering with polarizations
$\wedge\wedge\wedge\vee$.}
\label{uuud}
\end{center}
\end{figure}

As a preliminary, we evaluate the $\wedge\wedge\wedge\vee$
four-point tree with one leg off-shell, which will aid the construction 
of the loop integrand. For definiteness, let the $\vee$ helicity be
on leg 4 where legs 1234 are labeled counter-clockwise around the diagram,
as shown in Fig.~\ref{uuud}.
Also we omit the coupling factor $2g$ for each vertex. Then
\bea
A_{tree}^{\wedge\wedge\wedge\vee}&=&-{p^+_4\over p^+_1p^+_2p^+_3}
\left[{K^\wedge_{32}K^\wedge_{14}\over(p_2+p_3)^2}+
{K^\wedge_{43}K^\wedge_{21}\over(p_1+p_2)^2}\right]\\
\eea
Note that if $p_i^2=p_j^2=0$ we have the identity 
\bea K^\wedge_{ij}K^\vee_{ij}=-{1\over2}p^+_ip^+_j(p_i+p_j)^2\qquad{\rm for}~
p_i^2=p_j^2=0
\eea
With only one leg off-shell we can always write the denominators in terms of
on-shell momenta and exploit this identity. 
For example if only leg 4 is off-shell we have
\bea
A_{tree}^{\wedge\wedge\wedge\vee}&=&{1\over2}
\left[{p^+_4\over p^+_1}{K^\wedge_{14}\over K^\vee_{32}}+
{p^+_4\over p^+_3}{K^\wedge_{43}\over K^\vee_{21}}\right]=
{p^+_4\over2K^\vee_{32}K^\vee_{21}}\sum_i {K^\wedge_{4i}K^\vee_{i2}\over p^+_i}
=-{p^+_4p^+_2p_4^2\over4K^\vee_{32}K^\vee_{21}}\qquad {\rm for}~ 
p_1^2=p_2^2=p_3^2=0
\eea
which, of course, vanishes on-shell. Corresponding expressions when
other legs are off shell are obtained by always 
writing the denominators so they
involve the on shell legs only. We obtain
\bea
A_{tree}^{\wedge\wedge\wedge\vee}&=&
-{p^{+2}_4p^+_1p_3^2\over4p_3^+K^\vee_{14}K^\vee_{21}} 
\qquad {\rm for}~ p_1^2=p_2^2=p_4^2=0\\
A_{tree}^{\wedge\wedge\wedge\vee}&=&
-{p^{+2}_4p^+_3p_1^2\over4p_1^+K^\vee_{32}K^\vee_{43}} 
\qquad {\rm for}~ p_2^2=p_3^2=p_4^2=0\\
A_{tree}^{\wedge\wedge\wedge\vee}&=&
-{p^{+3}_4p_2^2\over4p_2^+K^\vee_{14}K^\vee_{43}} 
\qquad {\rm for}~ p_1^2=p_3^2=p_4^2=0
\eea
Thus the on-shell amplitudes for all like helicities 
and all but one like helicities vanish, a well-known result which applies
to the scattering of any number of gluons. 

For the case with all four gluons off-shell, the denominators
can't be factored so simply, but we can still get a reasonably
compact result.
\bea
A_{tree}^{\wedge\wedge\wedge\vee}&=&
-{p_4^+(K_{43}^\wedge K_{32}^\wedge p_1^2
+K_{14}^\wedge K_{43}^\wedge p_2^2+K_{21}^\wedge K_{14}^\wedge p_3^2
+K_{32}^\wedge K_{21}^\wedge p_4^2)\over p_1^+p_2^+p_3^+(p_1+p_2)^2
(p_2+p_3)^2}
\eea

\begin{figure}[ht]
\begin{center}
\psfrag{'1'}{$1$}
\psfrag{'2'}{$2$}
\psfrag{'3'}{$3$}
\psfrag{'4'}{$4$}
\includegraphics[width=3in]{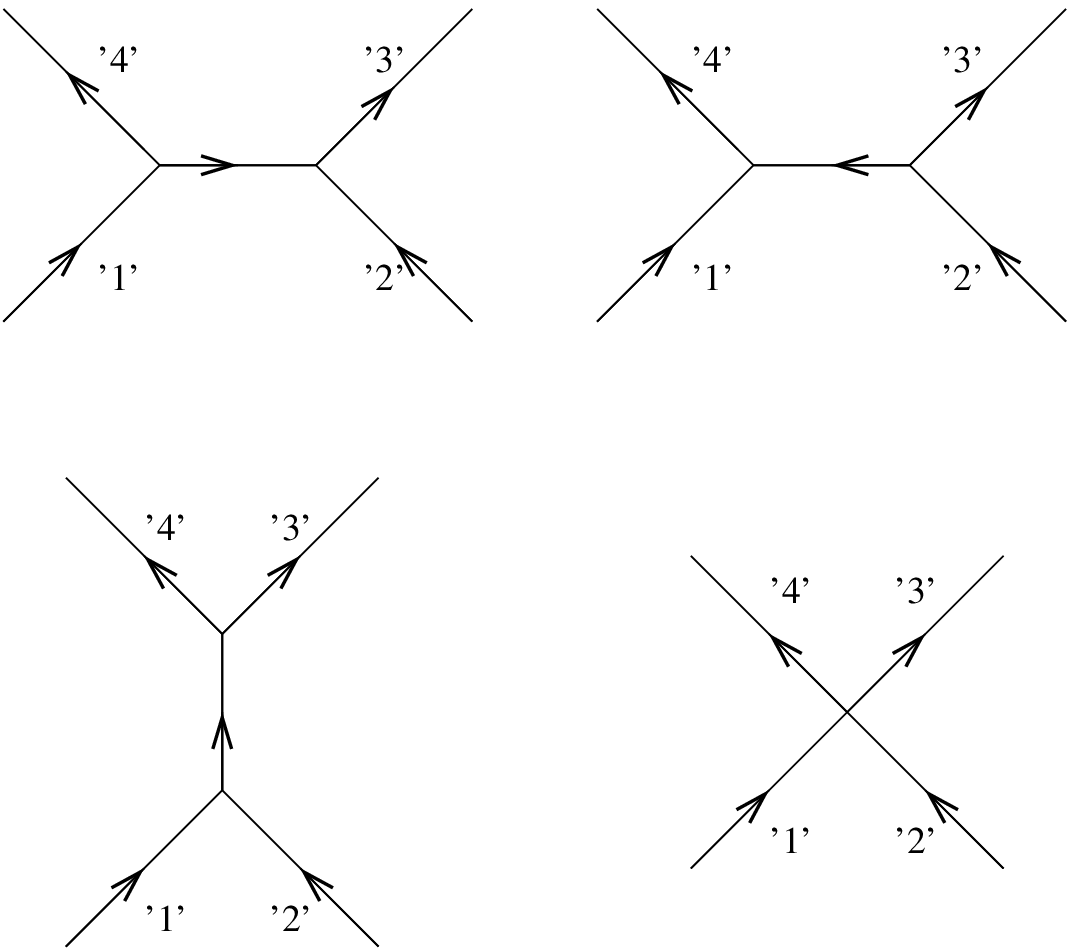}
\caption{Tree diagrams for the gluon scattering with polarizations
$\wedge\wedge\vee\vee$.}
\label{uudd}
\end{center}
\end{figure}

The only non-zero four point trees are those with two of each helicity.
The diagrams with adjacent like helicity are shown in Fig.~\ref{uudd}.
Applying the light-cone gauge rules to these diagrams gives
\bea
A_{tree}^{\wedge\wedge\vee\vee}&=&-{1\over(p_1^++p_4^+)^2}
\left[{p_1^+p_3^+\over p_2^+p_4^+}{K_{14}^\vee K_{32}^\wedge\over
(p_1+p_4)^2} +{p_2^+p_4^+\over p_1^+p_3^+}{K_{14}^\wedge K_{32}^\vee\over
(p_1+p_4)^2}+{p_1^+p_3^++p_2^+p_4^+\over 2}\right]\nonumber\\
&&\hskip.5in-{(p_1^++p_2^+)^2K_{21}^\wedge K_{43}^\vee
\over p_1^+p_2^+p_3^+p_4^+(p_1+p_2)^2}
\label{uuddamp}
\eea
With all legs on shell the right side can be dramatically simplified.
We shall need some further identities:
\bea
K_{14}^\wedge K_{32}^\vee+K_{14}^\vee K_{32}^\wedge
&=&-{p_1^+p_3^++p_2^+p_4^+\over 2}(p_1+p_4)^2+{(p_1^++p_4^+)^2\over2}
(p_1+p_2)^2\nonumber\\
&&+{p_1^++p_4^+\over2}
\left[p_1^+p_3^+(p_1^*-p_3^*)+p_2^+p_4^+(p_4^*-p_2^*)\right]\\
&\to&-{p_1^+p_3^++p_2^+p_4^+\over 2}(p_1+p_4)^2+{(p_1^++p_4^+)^2\over2}
(p_1+p_2)^2\quad ({\rm On~Shell})\nonumber
\eea
Here, when some legs are off-shell, we use the shorthand notation
$p_i^*\equiv p_i^2/p_i^+$ to simplify the writing. 
\bea
p_2^+p_4^+K_{14}^\wedge K_{32}^\vee-p_1^+p_3^+K_{14}^\vee K_{32}^\wedge
&=&-(p_1^++p_4^+)^2 K_{21}^\wedge K_{43}^\vee
+{(p_1^++p_4^+)}(p_1^+K_{32}^\wedge K_{43}^\vee-p_3^+K_{21}^\wedge K_{14}^\vee)
\nonumber\\
&=&-(p_1^++p_4^+)^2 K_{21}^\wedge K_{43}^\vee\nonumber\\
&&\qquad+{(p_1^++p_4^+)\over(p_1^++p_2^+)}
(p_1^+p_2^+K_{43}^\wedge K_{43}^\vee-2p_1^+p_3^+K_{21}^\wedge
K_{43}^\vee+p_3^+p_4^+K_{21}^\wedge K_{21}^\vee)\nonumber\\
&&\hskip-.75in =-{p_1^++p_4^+\over p_1^++p_2^+}\left[(p_1^+p_3^++p_2^+p_4^+) 
K_{21}^\wedge K_{43}^\vee
-p_1^+p_2^+K_{43}^\wedge K_{43}^\vee
-p_3^+p_4^+K_{21}^\wedge K_{21}^\vee\right]\nonumber\\
&&\hskip-.75in=-{p_1^++p_4^+\over p_1^++p_2^+}\bigg[(p_1^+p_3^++p_2^+p_4^+) 
K_{21}^\wedge K_{43}^\vee
+p_1^+p_2^+p_3^+p_4^+\bigg((p_1+p_2)^2\nonumber\\
&&\hskip1in-{1\over2}(p_1^++p_2^+)(p_1^*+p_2^*-p_3^*-p_4^*)\bigg)\bigg] \nonumber\\
&&\hskip-.75in\to-{p_1^++p_4^+\over p_1^++p_2^+}\left[(p_1^+p_3^++p_2^+p_4^+) 
K_{21}^\wedge K_{43}^\vee
+p_1^+p_2^+p_3^+p_4^+(p_1+p_2)^2\right] \quad ({\rm On~Shell})
\eea
We use these identities
to manipulate the first line to a form which shows no singularity
at $p_1^++p_4^+=0$ when all legs are on-shell:
\bea
A_{tree}^{\wedge\wedge\vee\vee}&=&{(p_1+p_2)^2\over2(p_1+p_4)^2}
-{K_{21}^\wedge K_{43}^\vee
\left[(p_1^++p_2^+)^2(p_1+p_4)^2-(p_1^+p_3^++p_2^+p_4^+)(p_1+p_2)^2\right]
\over p_1^+p_2^+p_3^+p_4^+(p_1+p_2)^2(p_1+p_4)^2}\nonumber\\
&& -{p_2^+p_4^+(p_1^*-p_3^*)+p_1^+p_3^+(p_4^*-p_2^*)
\over2(p_1^++p_4^+)(p_1+p_4)^2}
\eea
The quantity in square brackets in the numerator of the second term
can be related to
\bea
2(K_{21}^\wedge K_{43}^\vee+K_{21}^\vee K_{43}^\wedge)
&=&-(p_1^+p_3^++p_2^+p_4^+)(p_1+p_2)^2+{(p_1^++p_2^+)^2}
(p_1+p_4)^2\nonumber\\
&&+(p_1^++p_2^+)[p_1^+p_3^+(p_1^*-p_3^*)+p_2^+p_4^+(p_2^*-p_4^*)]
\eea
so it easily follows that
\bea
A_{tree}^{\wedge\wedge\vee\vee}&=&{(p_1+p_2)^2\over2(p_1+p_4)^2}
-{2(K_{21}^{\wedge2} K_{43}^{\vee2}+K_{21}^{\wedge}K_{21}^{\vee}
K_{43}^{\wedge} K_{43}^{\vee})
\over p_1^+p_2^+p_3^+p_4^+(p_1+p_2)^2(p_1+p_4)^2}
-{p_2^+p_4^+(p_1^*-p_3^*)+p_1^+p_3^+(p_4^*-p_2^*)
\over2(p_1^++p_4^+)(p_1+p_4)^2}\nonumber\\
&&+{K_{21}^\wedge K_{43}^\vee
(p_1^++p_2^+)[p_1^+p_3^+(p_1^*-p_3^*)+p_2^+p_4^+(p_2^*-p_4^*)]
\over p_1^+p_2^+p_3^+p_4^+(p_1+p_2)^2(p_1+p_4)^2}\\ 
&=&-{2K_{21}^{\wedge2} K_{43}^{\vee2}
\over p_1^+p_2^+p_3^+p_4^+(p_1+p_2)^2(p_1+p_4)^2}\nonumber\\
&&+{(p_1^++p_2^+)(p_1^*+p_2^*-p_3^*-p_4^*)\over2(p_1+p_4)^2}
-{p_2^+p_4^+(p_1^*-p_3^*)+p_1^+p_3^+(p_4^*-p_2^*)
\over2(p_1^++p_4^+)(p_1+p_4)^2}\\
&&+{(p_1^++p_2^+)^2(p_1^*+p_2^*)(p_3^*+p_4^*)
\over2(p_1+p_2)^2(p_1+p_4)^2}+{K_{21}^\wedge K_{43}^\vee
(p_1^++p_2^+)[p_1^+p_3^+(p_1^*-p_3^*)+p_2^+p_4^+(p_2^*-p_4^*)]
\over p_1^+p_2^+p_3^+p_4^+(p_1+p_2)^2(p_1+p_4)^2}\nonumber\\
&\to&-{2K_{21}^{\wedge2} K_{43}^{\vee2}
\over p_1^+p_2^+p_3^+p_4^+(p_1+p_2)^2(p_1+p_4)^2}\hskip2in{\rm (On~Shell)}\\
&=&{p_3^+p_4^+K_{12}^{\wedge4} 
\over 2p_1^+p_2^+K_{12}^\wedge K_{23}^\wedge K_{34}^\wedge K_{41}^\wedge}
\label{uuddtree}
\eea
which is essentially the Parke-Taylor form of the answer. The bispinor matrix
elements employed by them differ from our $K_{ij}$ by factors of $p^+$. 

\begin{figure}[ht]
\begin{center}
\psfrag{'1'}{$1$}
\psfrag{'2'}{$2$}
\psfrag{'3'}{$3$}
\psfrag{'4'}{$4$}
\includegraphics[width=4in]{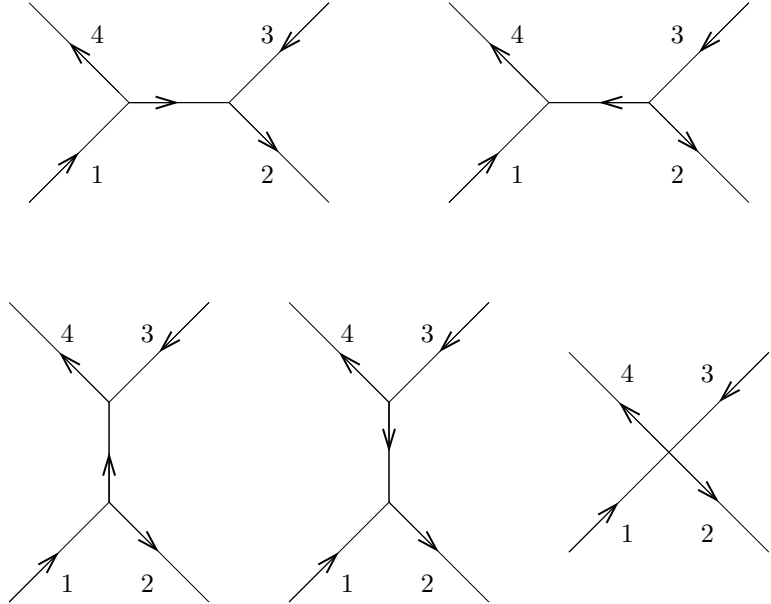}
\caption{Tree diagrams for gluon scattering with alternating helicity,
$\wedge\vee\wedge\vee$}
\label{udud}
\end{center}
\end{figure}

The other distinct helicity arrangement for four gluon
scattering is shown in Fig.~\ref{udud}. The light-cone gauge rules for
these diagrams give
\bea
A_{tree}^{\wedge\vee\wedge\vee}&=&-{1\over(p_1^++p_4^+)^2}
\left[{p_1^+p_2^+\over p_3^+p_4^+}{K_{14}^\vee K_{32}^\wedge\over
(p_1+p_4)^2} +{p_3^+p_4^+\over p_1^+p_2^+}{K_{14}^\wedge K_{32}^\vee\over
(p_1+p_4)^2}-{p_1^+p_2^++p_3^+p_4^+\over 2}\right]\nonumber\\
&&-{1\over(p_1^++p_2^+)^2}
\left[{p_1^+p_4^+\over p_2^+p_3^+}{K_{43}^\wedge K_{21}^\vee\over
(p_1+p_2)^2} +{p_2^+p_3^+\over p_1^+p_4^+}{K_{43}^\vee K_{21}^\wedge\over
(p_1+p_2)^2}-{p_1^+p_4^++p_2^+p_3^+\over 2}\right]
\label{ududamp}
\eea
where the quartic vertex contribution has been split between the
last terms in each of the square brackets. Notice that the second line on the
right side is obtained from the first line with the relabeling 
substitutions $1\to2\to3\to4\to1$ and $\wedge\to\vee\to\wedge$.
Furthermore the first line can be obtained from the first line on the
right of (\ref{uuddamp}) by interchanging $2\leftrightarrow3$ and
multiplying by the factor $-1$. Thus by inspection we immediately obtain 
the simplifications
\bea
A_{tree}^{\wedge\vee\wedge\vee}&=&-{p_{13}^2\over 2 p_{14}^2}
-{(p_1^+p_2^++p_3^+p_4^+)K_{31}^\wedge K_{42}^\vee\over p_1^+p_2^+p_3^+p_4^+
p_{14}^2}
-{p_{13}^2\over2p_{12}^2}
-{(p_2^+p_3^++p_1^+p_4^+)K_{13}^\wedge K_{42}^\vee\over p_1^+p_2^+p_3^+p_4^+
p_{12}^2}\nonumber\\
&&+{p_3^+p_4^+(p_1^*-p_2^*)+p_1^+p_2^+(p_4^*-p_3^*)\over
2p_{14}^2p_{14}^+}+{p_1^+p_4^+(p_2^*-p_3^*)+p_2^+p_3^+(p_1^*-p_4^*)\over
2p_{12}^2p_{12}^+}\nonumber\\
&=&-{p_{13}^2(p_{12}^2+p_{14}^2)\over2 p_{14}^2 p_{12}^2}
-{2K_{31}^\wedge K_{42}^\vee [K_{31}^\wedge K_{42}^\vee
+K_{31}^\vee K_{42}^\wedge]\over p_1^+p_2^+p_3^+p_4^+
p_{14}^2 p_{12}^2}\nonumber\\
&&-K_{31}^\wedge K_{42}^\vee{
p_3^+p_4^+(p_1^2+p_2^2)+p_1^+p_2^+(p_3^2+p_4^2)
-p_2^+p_3^+(p_1^2+p_4^2)-p_1^+p_4^+(p_2^2+p_3^2)\over p_1^+p_2^+p_3^+p_4^+
p_{14}^2 p_{12}^2}\nonumber\\
&&+{p_3^+p_4^+(p_1^*-p_2^*)+p_1^+p_2^+(p_4^*-p_3^*)\over
2p_{14}^2p_{14}^+}+{p_1^+p_4^+(p_2^*-p_3^*)+p_2^+p_3^+(p_1^*-p_4^*)\over
2p_{12}^2p_{12}^+}\nonumber\\
&=&
-{2K_{31}^{\wedge2} K_{42}^{\vee2}\over p_1^+p_2^+p_3^+p_4^+
p_{12}^2p_{14}^2}-{p_{13}^+p_{24}^+(p_1^*+p_3^*)(p_2^*+p_4^*)\over
2p_{12}^2p_{14}^2}+{p_{13}^2[p_4^+p_2^*+p_2^+p_4^*+p_3^+p_1^*+p_1^+p_3^*]
\over2p_{12}^2p_{14}^2}\nonumber\\
&&-K_{31}^\wedge K_{42}^\vee{
p_3^+p_4^+(p_1^2+p_2^2)+p_1^+p_2^+(p_3^2+p_4^2)
-p_2^+p_3^+(p_1^2+p_4^2)-p_1^+p_4^+(p_2^2+p_3^2)\over p_1^+p_2^+p_3^+p_4^+
p_{14}^2 p_{12}^2}\nonumber\\
&&+{p_3^+p_4^+(p_1^*-p_2^*)+p_1^+p_2^+(p_4^*-p_3^*)\over
2p_{14}^2p_{14}^+}+{p_1^+p_4^+(p_2^*-p_3^*)+p_2^+p_3^+(p_1^*-p_4^*)\over
2p_{12}^2p_{12}^+}\nonumber\\
&\to&
-{2K_{31}^{\wedge2} K_{42}^{\vee2}\over p_1^+p_2^+p_3^+p_4^+
p_{12}^2p_{14}^2} ,\hskip1.5in{\rm (On~Shell)}\\
&=&{p_2^+p_4^+K_{13}^{\wedge4} 
\over 2p_1^+p_3^+K_{12}^\wedge K_{23}^\wedge K_{34}^\wedge K_{41}^\wedge}
\eea
where we have used the shorthand notation $p_{ij}=p_i+p_j$. Again we
note the characteristic Parke-Taylor form in the last line.

\section{Gluon Self-Energy}
In order to acquaint the reader with some of the novelties
of calculations using the $\delta$ regulator of ultraviolet
divergences, we calculate the
gluon self-energy diagrams in complete detail,
with an explicit separation of all divergences and 
Lorentz-violating artifacts. Recall that the choice of
light-cone gauge has sacrificed manifest Lorentz invariance,
leaving only the light-cone Galilei subgroup of the
Poincar\'e group as a manifest symmetry. In addition, discretization of
$p^+$ provides an infrared regulator that breaks the
Lorentz symmetry generated by $M^{+-}$, under which $p^+$ and $p^-$ scale
oppositely. Another novelty in the calculation comes from 
our use of dual momentum variables: for an $N$ point
amplitude there are $N$ independent dual momenta
but only $N-1$ independent actual momenta (because of momentum conservation).
In the bare Feynman rules this redundancy is taken care of by
a new symmetry under the translation of all dual momenta
by the same amount. The $\delta$ regulator inserts
a factor $e^{-\delta\sum{\bfs q_i}^2}$, where ${\bfs q}_i$ are
the dual momenta assigned to the loops. With $\delta>0$
the dual momentum translation invariance is explicitly broken,
and so is the Galilei boost part of the residual light-cone
Galilei invariance. The upshot of all this \cite{thornscalar} 
is that in the
presence of nonzero $\delta$ and discrete $p^+$,
manifest Poincar\'e invariance is broken to translation invariance
in the $+$ and transverse directions
and the $O(2)$ rotation invariance in the transverse plane, as
well as the conservation of discrete $p^+$. We demand that
the full Poincar\'e symmetry is restored as $\delta\to0$ and
$p^+$ becomes continuous, and we shall determine the
counterterms necessary to achieve this to one-loop order.

We call the bare gluon self-energy $\Pi^{ij}$, and stress that
it includes contributions arising from the elimination of
the gauge field components $A^{\pm}$ from the formalism.
We continue to use the helicity 
basis for the gluon polarization $1,2$: $\wedge=(1+i2)/\sqrt2$,
$\vee=(1-i2)/\sqrt2$. 
It is convenient to introduce the Schwinger representation for each internal 
propagator
\bea
{e^{-\delta{\bfs q}^2}\over (q-k_0)^2(q-k_1)^2}=
\int dT_1dT_2e^{-T_1(q-k_0)^2-T_2(q-k_1)^2-\delta{\bfs q}^2}
\eea
where we have used dual momenta $k_0,k_1$ with
the momentum carried by the propagator $p=k_1-k_0$, and we take $k_0^+=0$.
The standard light-cone evaluation of the $q^-$ integration using the
calculus of residues is equivalent
to the insertion $\pi\delta((T_{1}+T_2)q^+-T_2p^+)$ in the integral over 
the remaining components of momentum. The integral over transverse momentum 
is easily done by completing the square in the exponent. We leave $q^+=lm$,
$l=1,2,\ldots, M=p^+/m$,
discretized to regulate the $q^+=0$ singularities.
\begin{figure}[ht]
\psfrag{'q'}{$q$}
\psfrag{'k1'}{$k_0$}
\psfrag{'k2'}{$k_1$}
\begin{center}
\includegraphics[height=3cm]{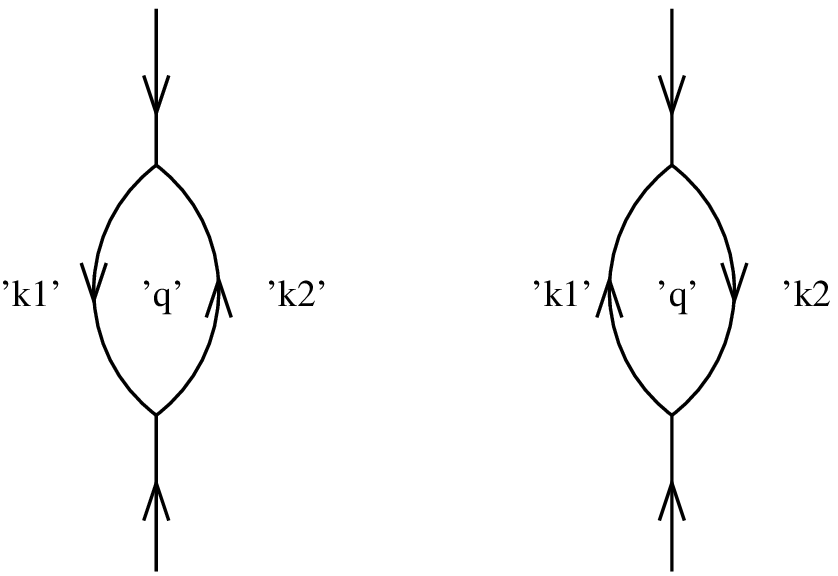}\\
\includegraphics[height=3cm]{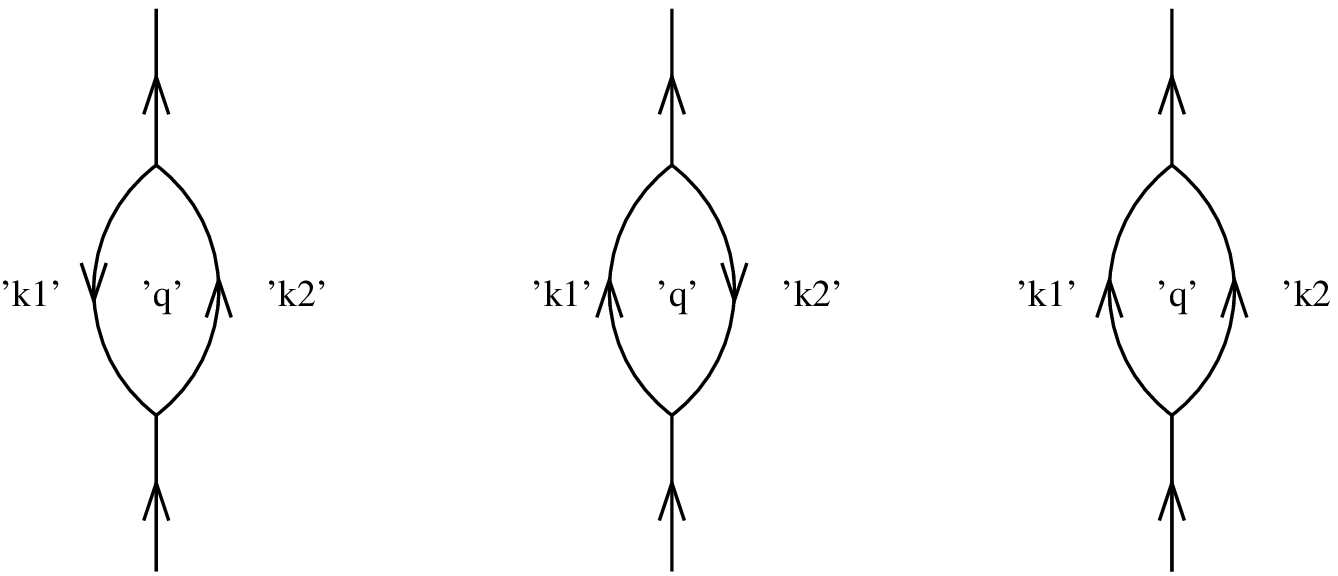}
\caption{The self energy diagrams for $\Pi^{\wedge\wedge}$ (top line)
and $\Pi^{\wedge\vee}$ (bottom line).}
\label{segraphs}
\end{center}
\end{figure}
First consider the like helicity component $\Pi^{\wedge\wedge}$.
\bea
\Pi^{\wedge\wedge}&=&{g^2N_c\over2\pi^2}\int_0^\infty dT\int_0^1dx 
{\delta^2[x{k_0^\vee}+(1-x){k}_1^{\vee}]^2\over
(T+\delta)^3}\nonumber\\&& 
\exp\left\{-Tx(1-x)p^2
-{\delta T\over T+\delta}(x{\boldsymbol k}_0+(1-x){\boldsymbol k}_1)^2
\right\}\\
&\to& {g^2N_c\over4\pi^2}\int_0^1dx 
{[x{k_0^\wedge}+(1-x){k}_1^{\wedge}]^2}={g^2N_c\over12\pi^2} 
[{k_0^{\wedge2}}+{k}_1^{\wedge2}+k_0^\wedge k_1^\wedge]
\eea
where we have included the factor of $N_c$ arising from our substitution
rule giving the color dependence of the large $N_c$ limit. In this expression,
it was safe to take $q^+$ continuous since there are no $q^+=0$
singularities in the integrand. We have also changed variables
from Schwinger parameters to $T=T_1+T_2$, $x=T_1/(T_1+T_2)$.
Lorentz invariance would
imply that $\Pi^{\wedge\wedge}=0$ on-shell, otherwise a single isolated
gluon could flip its helicity. Indeed, the Galilei boost invariance alone
would imply that it vanishes. But the $\delta$ regulator breaks
this invariance. We see by inspection that the whole $T$ integration
region a finite distance away from the origin vanishes as
$\delta\to0$, but the region with $T=O(\delta)$ survives the
limit due to the $(T+\delta)^{-3}$ behavior of the integrand.
The result is nevertheless finite and a quadratic polynomial
in the dual momenta. Note that in addition to violating
Lorentz invariance the result also violates the dual momentum translation
symmetry. We therefore must introduce a counterterm that exactly
cancels this result:
\bea
\Pi^{\wedge\wedge}_{\rm TOT}\equiv
\Pi^{\wedge\wedge}+\Pi_{\rm C.T.}^{\wedge\wedge}=0.
\eea

We now briefly discuss how this counterterm is incorporated in the
worldsheet description. The dual momenta ${\bfs k}_0,{\bfs k}_1$
are the boundary values of the worldsheet field ${\bfs q}(\sigma,\tau)$.
We can write the contribution of the counterterm to the
worldsheet path integral as
\bea
{-T\over2p^+}\Pi^{\wedge\wedge}_{\rm C.T.}&=&{T\over2p^+}{g^2N_c\over12\pi^2} 
\left[{3\over2}{k_0^{\wedge2}}+{3\over2}{k}_1^{\wedge2}
-{1\over2}p^{\wedge2}\right]\nonumber\\
&=&{g^2N_c\over16\pi^2}\int d\tau{{q^{\wedge2}(0)}+{q}^{\wedge2}(p^+)
\over p^+}
-{g^2N_c\over48\pi^2}\int d\tau d\sigma \left({\partial q^\wedge\over
\partial\sigma}\right)^2
\eea
We see that the counter term can be associated with new terms in the
worldsheet action, in a fashion similar to
that in \cite{thornscalar} The first term is a boundary term and the second
is a bulk term. In order to ensure that the term enters only
for like helicities, additional factors of Grassmann variables
${\bar S}^\vee(\sigma) S^\vee(\sigma)$ must also be included 
(see \cite{thornsheet}).

Next we turn to the helicity conserving contributions to $\Pi$.
\bea
\Pi^{\wedge\vee}&=&\Pi^{\vee\wedge}
={g^2N_c\over4\pi^2}\sum_{q^+}\int_0^\infty{dT}
\int_0^1 dx\delta(q^+-(1-x)p^+)\left[{1\over(T+\delta)^2}
+{\delta^2[x{\bfs k}_0+(1-x){\bfs k}_1]^2\over
(T+\delta)^3}\right]\nonumber\\
&&\qquad\left(1+{1\over x}
+{1\over (1-x)^2}\right)
\exp\left\{-Tx(1-x)p^2
-\delta T{(x{\boldsymbol k}_0+(1-x){\boldsymbol k}_1)^2
\over(T+\delta)}\right\}\nonumber\\
&=&
{g^2N_c\over4\pi^2}{1\over p^+}\sum_{0<q^+<p^+}\int_0^\infty dT
\left[{1\over (T+\delta)^2}
+{\delta^2[x{\bfs k}_0+(1-x){\bfs k}_1]^2\over
(T+\delta)^3}\right]\nonumber\\
&& \left(1+{1\over x^2}
+{1\over (1-x)^2}\right)
\exp\left\{-Tx(1-x)p^2
-{\delta T\over T+\delta}(x{\boldsymbol k}_0+(1-x){\boldsymbol k}_1)^2
\right\}\bigg|_{x=1-q^+/p^+}
\eea
The quadratic divergence in $\Pi^{\wedge\vee}$ can be 
simply extracted with an integration
by parts. We observe that
\bea
\left[{1\over (T+\delta)^2}
+{\delta^2[x{\bfs k}_0+(1-x){\bfs k}_1]^2\over
(T+\delta)^3}\right]\exp\left\{-{\delta T\over T+\delta}
(x{\boldsymbol k}_0+(1-x){\boldsymbol k}_1)^2
\right\}&=&\nonumber\\
&&\hskip-1.5in -{\partial\over\partial T}{1\over T+\delta}
\exp\left\{-{\delta T\over T+\delta}
(x{\boldsymbol k}_0+(1-x){\boldsymbol k}_1)^2
\right\}
\eea
So we can rewrite the self energy as
\bea
\Pi^{\wedge\vee}&=&-{g^2N_c\over4\pi^2}p^2
\sum_{q^+}{1\over p^+}\left({q^+(p^+-q^+)\over p^{+2}}+{p^+-q^+\over q^+}
+{q^+\over p^+-q^+}\right) I( H\delta) \nonumber\\&&
+{g^2N_c\over4\pi^2}{1\over\delta}
\sum_{q^+}{1\over p^+}\left(1+{p^{+2}\over q^{+2}}
+{p^{+2}\over (p^+-q^+)^{2}}\right)\\
H&\equiv&x(1-x)p^2,\qquad\qquad x=1-{q^+\over p^+}\\
I(H\delta)&\equiv&\int_0^\infty {e^{-H\delta u-\delta u(x{\bfs k}_0
+(1-x){\bfs k}_1)^2/(1+u)
}du\over 1+u}\quad{}_{\widetilde{\delta\to0}} 
\quad -\gamma-\ln\{H\delta\}
\eea
where $\gamma=-\Gamma^\prime(1)/\Gamma(1)$ is Euler's constant.
Note that the dual momentum translation invariance is restored in
this quantity as $\delta\to0$, apparently for accidental reasons.
Clearly the $q^+$ sums diverge when $q^+$ becomes continuous.
These divergences are spurious artifacts of the light-cone gauge
and have nothing to do with the usual ultraviolet divergences of the
gauge theory. They must cancel in physical quantities without invoking
renormalization or counterterms. In our approach the $q^+$ sum
just corresponds to integration over the location on the
worldsheet of the boundary representing the loop. On the
worldsheet lattice this location is an integer $l$ with 
$x=l/M$ and $M$ is the discretized total plus momentum
entering the self energy: $p^+=mM$. In the above
formulas, $\sum_{q^+}$ means $m\sum_{l=1}^{M-1}$. The discreteness of
$p^+$ regulates the endpoint $x$ integral divergences. 

Let us discuss first the fate of the quadratic $1/\delta$ divergence,
which for discrete $p^+$ reads, with $q^+=lm$:
\bea
{g^2N_c\over4\pi^2}{1\over M\delta }\sum_{l=1}^{M-1}
\left(1+{M^2\over l^2}+{M^2\over(M-l)^2}\right)
\sim {g^2N_c\over4\pi^2}{1\over\delta}
\left({\pi^2\over3}M-1+O\left({1\over M}\right)\right)
\eea
where the right side indicates the large $M$ behavior of the
sums. The term linear in $M=p^+/m$ cannot be canceled by a
gluon self mass, because it is linear in $p^+$. However, 
precisely because it is linear in $p^+$, it represents a
constant $-g^2N_cM/(24p^+\delta)=-g^2N_c/(24m\delta)$ 
added to the energy $p^-={\bfs p}^2/2p^+$ of each gluon. 
Of course the number of gluons is not fixed as a function of time
so we can't say that this constant is unobservable. If there are
$n$ gluons present at a given time, a constant $e_0$ added to each
gluon energy would add $ne_0$ to the total energy. In the worldsheet picture
the number of gluons changes whenever an internal boundary terminates
or a new one originates. And the contribution of each gluon
to the worldsheet action would be $-e_0t$ where t is the time the gluon exists.
Thus a constant added to $p^-$
can be interpreted as
energy associated with the boundary of the worldsheet representing
that gluon. The corresponding contribution to the worldsheet
action is then $-e_0L/2$ where $L$ is the sum of all the lengths
of all the boundaries in the worldsheet. (Note that
an internal boundary representing a loop has total length $2T$
where $T$ is the time the boundary exists.)
In other words $e_0/2$ contributes like a boundary cosmological constant. If we
start with a nonzero boundary cosmological constant $\lambda_b$ in
zeroth order, we can tune its value to cancel the
linear terms in $p^+$ generated by loop effects. Its lowest order value is 
then
\bea
\lambda_b=+{g^2N_c\over48m\delta}
\eea
After this cancellation,
there is left behind a constant which can be canceled by a gluon mass
counterterm $\delta\mu^2$. So to this order
\bea
\delta\mu^2={g^2N_c\over4\pi^2\delta}
\eea 
Of course, the gluon mass is zero in tree approximation,
but since loop corrections generate a gluon mass, the tree value
must be non-zero and adjusted to cancel the loop contributions
order by order in perturbation theory. A nonzero mass at tree level 
violates gauge invariance, which means a violation of Lorentz invariance
in the completely fixed lightcone gauge. So an alternative prescription
is: in lightcone gauge, allow a nonzero gluon mass $\mu_0^2$ 
as an input parameter,
and calculate physical quantities as functions of this parameter.
Finally, choose a value of this parameter that restores Lorentz invariance.
Note that to one loop, $\mu^2=0$ requires a tachyonic gluon mass:
$\mu_0^2=-\delta\mu^2$. A gluon mass is introduced into the
worldsheet formalism for gauge theory 
exactly as in the scalar case \cite{thornscalar}.

Next we turn to the logarithmic divergences in the self energy.
Call $p=q^\prime-q$ and remember that
$x=q^+/p^+$. Including the above mentioned counterterms, we
then find 
\bea
\Pi^{\wedge\vee}_{\rm TOT}&=&\Pi^{\wedge\vee}-2p^+\lambda_b+\delta\mu^2
\nonumber\\ 
&=& {g^2N_c\over4\pi^2}p^2
\sum_{q^+}\left({x(1-x)-2\over p^+}+{1\over q^+}
+{1\over p^+-q^+}\right)\ln\left\{x(1-x)p^2\delta e^\gamma\right\}\\
&\sim&
{g^2N_c\over4\pi^2}p^2
\left(\sum_{q^+}\left[{1\over q^+}
+{1\over p^+-q^+}\right]\ln\left\{{q^+(p^+-q^+)\over p^{+2}}
p^2\delta e^\gamma\right\}
-{11\over6}\ln(p^2\delta e^\gamma)+{67\over18}\right)
\eea
Note that this quantity is negative by virtue of the (divergent) $q^+$ sums,
in accordance with the requirements of unitarity.  
The divergent $p^+$ dependent coefficient of $\ln\delta$ is characteristic
of light-cone gauge. We shall find that these unusual terms cancel
against corresponding terms coming from triangle and box diagrams.
To simplify future equations, we give the anomalous quantity in 
parentheses a name:
\bea
{\cal A}(p^2,p^+)\equiv\sum_{q^+}\left[{1\over q^+}
+{1\over p^+-q^+}\right]\ln\left\{x(1-x)p^2\delta e^\gamma\right\}.
\eea
We shall find this quantity occurring in vertex calculations. Then
to summarize this section:
\bea
\Pi_{\rm TOT}^{\wedge\vee}&=& 
{g^2N_c\over4\pi^2}p^2\left[{\cal A}(p^2,p^+)
-{11\over6}\ln\{p^2\delta e^\gamma\}+{67\over18}\right]\\
\Pi_{\rm TOT}^{\wedge\wedge}&=&\Pi_{\rm TOT}^{\vee\vee}=0        
\eea
\section{Cubic Vertex Function}
\begin{figure}[ht]
\psfrag{'p1+'}{$p_1^+$}
\psfrag{'p2+'}{$p_2^+$}
\psfrag{'A'}{$A$}
\psfrag{'B'}{$B$}
\psfrag{'k0'}{$k_0$}
\psfrag{'k1'}{$k_1$}
\psfrag{'k2'}{$k_2$}
\begin{center}
\includegraphics[width=6cm]{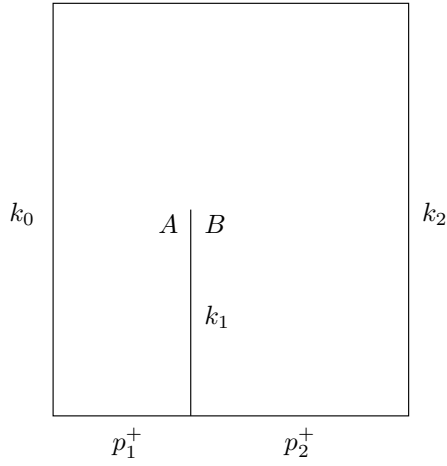}
\caption{The lightcone worldsheet for the cubic vertex}
\label{cubicws}
\end{center}
\end{figure}

We shall not include calculational details for the one loop
corrections to the cubic vertex function. They can be found in
\cite{thornlcnotes}. Instead we present the final answers
for the vertex corrections with two on-shell gluons.
\begin{figure}[ht]
\psfrag{'q'}{$q$}
\psfrag{'k0'}{$k_0$}
\psfrag{'k1'}{$k_1$}
\psfrag{'k2'}{$k_2$}
\begin{center}
\includegraphics[width=12cm]{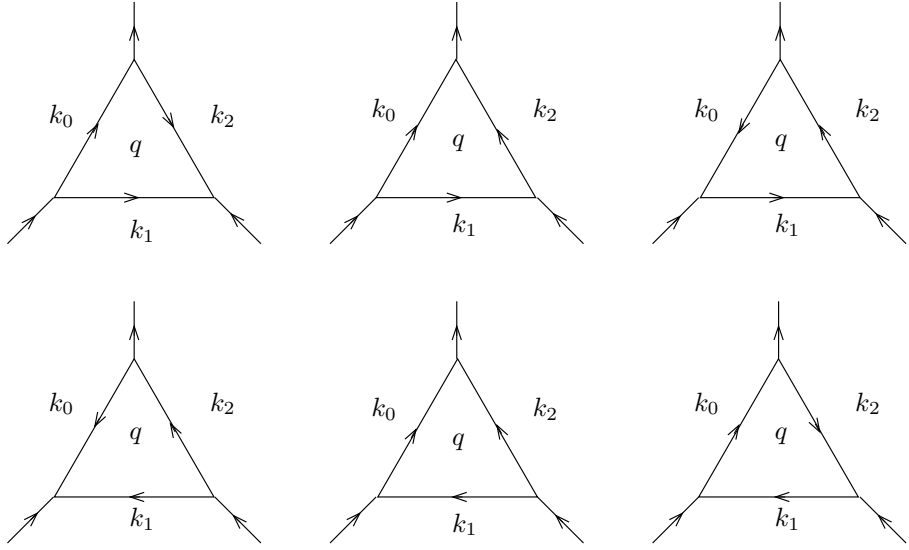}
\caption{The triangle diagrams contributing to $\Gamma^{\wedge\wedge\vee}$.}
\label{uudgraphs}
\end{center}
\end{figure}
\begin{figure}[ht]
\psfrag{'q'}{$q$}
\psfrag{'k0'}{$k_0$}
\psfrag{'k1'}{$k_1$}
\psfrag{'k2'}{$k_2$}
\begin{center}
\includegraphics[width=14cm]{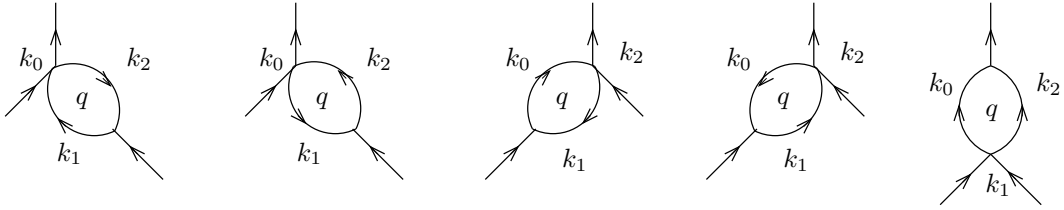}
\caption{The swordfish diagrams contributing to $\Gamma^{\wedge\wedge\vee}$.}
\label{uudsfgraphs}
\end{center}
\end{figure}
We put the 
combination of swordfish and triangle diagrams 
(see Figs.~\ref{uudgraphs},\ref{uudsfgraphs}) 
with two like-helicities
and two legs on-shell in the form
\bea
\Gamma_{\rm 1~loop}=-{(g\sqrt{N_c})^3\over12\pi^2}\sum_i k_i
-{g^2N_c\over8\pi^2}
\Gamma_{\rm tree}\left({70\over9}
-{11\over3}\ln(\delta p_o^2e^{\gamma})+S\right)
+\alpha{(g\sqrt{N_c})^3\over12\pi^2}{K\over p_o^+}
\label{cubiccorrection}
\eea
where the vectors $k_i,K$ carry the polarization of the two like-
helicity gluons, $p_o$ is the four-momentum of the off-shell gluon,
$\alpha=1$ when the on-shell gluons have like-helicity,
and $\alpha=0$ otherwise. Finally $S$
is an infrared sensitive term that depends on the location of
the off-shell gluon, but not on any of the gluon helicities. 
In the
case $p_1^+, p_2^+>0$, we denote by $S_i^{q^+}(p_1,p_2)$ the
value of $S$ when leg $i$ is off-shell, and with loop momentum chosen
so that $q^+$  is the longitudinal momentum of the
internal line joining leg $1$ to leg $3$, satisfying $0<q^+<p_{12}^+$.
Then, \bea
S_1^{q^+}(p_1,p_2)&=&\sum_{q^+<p_1^+}  
\bigg\{
\bigg[{2\over q^+}+{1\over p_1^++p_2^+-q^+}+{1\over p_1^+-q^+}
\bigg]\left(\ln(\delta p_1^2e^{\gamma})
+\ln{q^+\over p_1^+}\right)\nonumber\\
&&+\bigg[{2\over p_1^+-q^+}-{1\over p_1^++p_2^+-q^+}+{1\over q^+}
\bigg]\ln{p_1^+-q^+\over p_1^+}\bigg\}\nonumber\\
&&+\sum_{q^+>p_1^+} \bigg\{
\bigg[{1\over q^+}+{2\over p_1^++p_2^+-q^+}+{1\over q^+-p_1^+}
\bigg]\left(\ln(\delta p_1^2e^{\gamma})
+\ln{p_1^++p_2^+-q^+\over p_2^+}\right)
\nonumber\\
&&+\sum_{q^+\neq p_1^+}\bigg[{1\over q^+}+{2\over p_1^++p_2^+-q^+}
+{1\over q^+-p_1^+}
\bigg]\ln{p_1^++p_2^+-q^+\over p_1^++p_2^+}\\
S_2^{q^+}(p_1,p_2)&=&\sum_{q^+\neq p_1^+}
\left[{2\over q^+}+{1\over p_1^++p_2^+-q^+}+{1\over p_1^+-q^+}\right]
\ln{q^+\over p_1^++p_2^+}\nonumber\\
&&+\sum_{q^+<p_1^+}  
\bigg\{\bigg[{2\over q^+}+{1\over p_1^++p_2^+-q^+}+{1\over p_1^+-q^+}
\bigg]\left(\ln(\delta p_2^2e^{\gamma})+\ln{q^+\over p_1^+}\right)
\bigg\}\nonumber\\
&&+\sum_{q^+>p_1^+}\bigg\{
\bigg[{1\over q^+}+{2\over p_1^++p_2^+-q^+}+{1\over q^+-p_1^+}
\bigg]\left(\ln(\delta p_2^2e^{\gamma})
+\ln{p_1^++p_2^+-q^+\over p_2^+}\right)
\nonumber\\
&&\qquad\qquad
+\bigg[{2\over q^+-p_1^+}+{1\over p_1^++p_2^+-q^+}-{1\over q^+}
\bigg]
\ln{q^+-p_1^+\over p_2^+}\bigg\}\\
S_3^{q^+}(p_1,p_2)&=&\sum_{q^+<p_1^+}  
\bigg\{
\bigg[{2\over q^+}+{1\over p_1^++p_2^+-q^+}+{1\over p_1^+-q^+}
\bigg]\left(\ln(\delta p_{12}^2e^{\gamma})
+\ln{q^+\over p_1^++p_2^+}\right)\nonumber\\
&&+\bigg[{1\over q^+}+{2\over p_1^++p_2^+-q^+}+{1\over q^+-p_1^+}
\bigg]\ln{p_1^++p_2^+-q^+\over p_1^++p_2^+}\nonumber\\
&&+\bigg[{2\over p_1^+-q^+}-{1\over p_1^++p_2^+-q^+}+{1\over q^+}
\bigg]\ln{p_1^+-q^+\over p_1^+}\bigg\}\nonumber\\
&&+\sum_{q^+>p_1^+} \bigg\{
\bigg[{1\over q^+}+{2\over p_1^++p_2^+-q^+}+{1\over q^+-p_1^+}
\bigg]\left(\ln(\delta p_{12}^2e^{\gamma})
+\ln{p_1^++p_2^+-q^+\over p_1^++p_2^+}\right)
\nonumber\\
&&+\bigg[{2\over q^+}+{1\over p_1^++p_2^+-q^+}+{1\over p_1^+-q^+}
\bigg]\ln{q^+\over p_1^++p_2^+}
\nonumber\\ &&
+\bigg[{2\over q^+-p_1^+}+{1\over p_1^++p_2^+-q^+}-{1\over q^+}
\bigg]\ln{q^+-p_1^+\over p_2^+}\bigg\}
\eea
The first term on the right of (\ref{cubiccorrection}) 
must be canceled by a counterterm,
since it violates Lorentz invariance. The required counterterm
has the form
\bea
\Gamma_{\rm C.T.}^{\wedge\wedge\vee}=
+{(g\sqrt{N_c})^3\over12\pi^2}(k_0^\wedge+k_1^\wedge+k_2^\wedge)
\eea
which is completely symmetric in the three legs. We now show how
it can be described in the worldsheet formalism. The lightcone worldsheet
for the cubic vertex is a rectangle with a cut represented by a solid
line as in Fig.~\ref{cubicws}. 
Each dual momentum ${\bfs k}_i$ is
the boundary value of ${\bfs q}(\sigma,\tau)$ on one of the
three boundaries: the left side of the rectangle, the cut in the middle,
and the right side of the rectangle. For definiteness we assign these 
boundaries the label $i=0,1,2$, so ${\bfs k}_i$ is the boundary
value at boundary $i$. The cubic vertex is characterized by the point
on the worldsheet where the cut terminates. The value of
${\bfs q}$ on the cut is ${\bfs k}_1$ and by continuity
it has this value at the termination point. Thus we
can write $q^\wedge(A)=k^\wedge_1$.  That  is  ${k}_1^\wedge$
is locally associated with the vertex but $k_0^\wedge$ and $k_2^\wedge$
are not. However an insertion of $\partial q^\wedge/\partial\sigma$
into the worldsheet path integral of a single gluon produces
the factor \cite{thornsheet}
\bea
\left\langle{\partial q^\wedge\over\partial\sigma}\right\rangle
={\Delta q^\wedge\over p^+}
\eea
where $\Delta q^\wedge$ is the difference of the boundary values at the
two boundaries of the gluon worldsheet. Thus we have
\bea
\left\langle{\partial q^\wedge\over\partial\sigma}\right\rangle_{01}&=&
{k_1^\wedge-k_0^\wedge\over p_1^+}\\
\left\langle{\partial q^\wedge\over\partial\sigma}\right\rangle_{12}&=&
{k_2^\wedge-k_1^\wedge\over p_2^+}
\eea
where the $ij$ subscript means that the insertion is somewhere
between boundaries $i$ and $j$, $p_1^+$ the $+$ momentum carried
by the left gluon is the width of the left gluon strip, and
$p_2^+$ is the width of the right gluon strip. Since it doesn't matter
exactly where the insertion occurs we are free to make it
arbitrarily close to the end of the internal boundary, say the
points $A,B$ in the figure. Thus we can 
write the counterterm
\bea
\Gamma_{\rm C.T.}^{\wedge\wedge\vee}=+{(g\sqrt{N_c})^3\over12\pi^2}
\left[3q^\wedge(A)
+p_2^+\left\langle{\partial q^\wedge\over\partial\sigma}(B)\right\rangle
-p_1^+\left\langle{\partial q^\wedge\over\partial\sigma}(A)\right\rangle\right]
\eea
This is still not quite a local worldsheet modification because of the
factors $p_i^+$. But as shown in \cite{thornsheet} these factors can
be locally reproduced by inserting worldsheet ghost fields near
the interaction point. Thus the required counterterm has a local worldsheet
representation. Including the counterterm we then have for the 
complete vertex function
\bea
\Gamma_{\rm 1~loop}+\Gamma_{\rm C.T.}
=-{g^2N_c\over8\pi^2}
\Gamma_{\rm tree}\left({70\over9}
-{11\over3}\ln(\delta p_o^2e^{\gamma})+S\right)
+\alpha{(g\sqrt{N_c})^3\over12\pi^2}{K\over p_o^+}
\label{cubiccomplete}
\eea

\begin{figure}[ht]
\psfrag{'q'}{$q$}
\psfrag{'k0'}{$k_0$}
\psfrag{'k1'}{$k_1$}
\psfrag{'k2'}{$k_2$}
\begin{center}
\includegraphics[width=8cm]{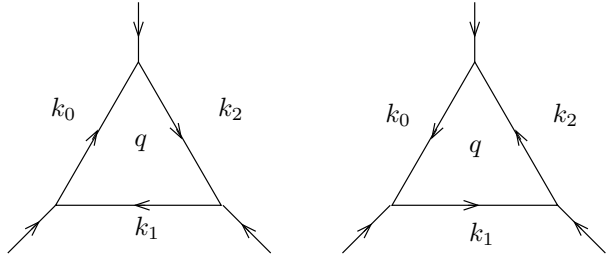}
\caption{The triangle diagrams contributing to $\Gamma^{\wedge\wedge\wedge}$.}
\label{uuugraphs}
\end{center}
\end{figure}

In addition to these corrections to the tree level cubic vertex, the
triangle diagram with three like-helicities (see Fig.~\ref{uuugraphs}) 
is non-zero, and it is given,
for the case of two on-shell legs, by
\bea
\Gamma_\triangle^{\wedge\wedge\wedge}&=&-{(g\sqrt{N_c})^3\over6\pi^2}
{K^{\wedge3}
\over p_1^+p_2^+p_3^+ p_o^2}\\
\Gamma_\triangle^{\vee\vee\vee}&=&-{(g\sqrt{N_c})^3\over6\pi^2}{K^{\vee3}
\over p_1^+p_2^+p_3^+ p_o^2}
\eea
where $p_o$ is the momentum of the off-shell gluon.

\section{Reduction of Box Diagrams}
\begin{figure}[ht]
\psfrag{'='}{$=$}
\psfrag{'-'}{$-$}
\psfrag{'triangle'}{\large Triangle-like}
\psfrag{'+'}{$\hskip -.25in +$}
\psfrag{'q'}{$q$}
\psfrag{'k0'}{$k_0$}
\psfrag{'k1'}{$k_1$}
\psfrag{'k2'}{$k_2$}
\psfrag{'k3'}{$k_3$}
\begin{center}
\includegraphics[width=5in]{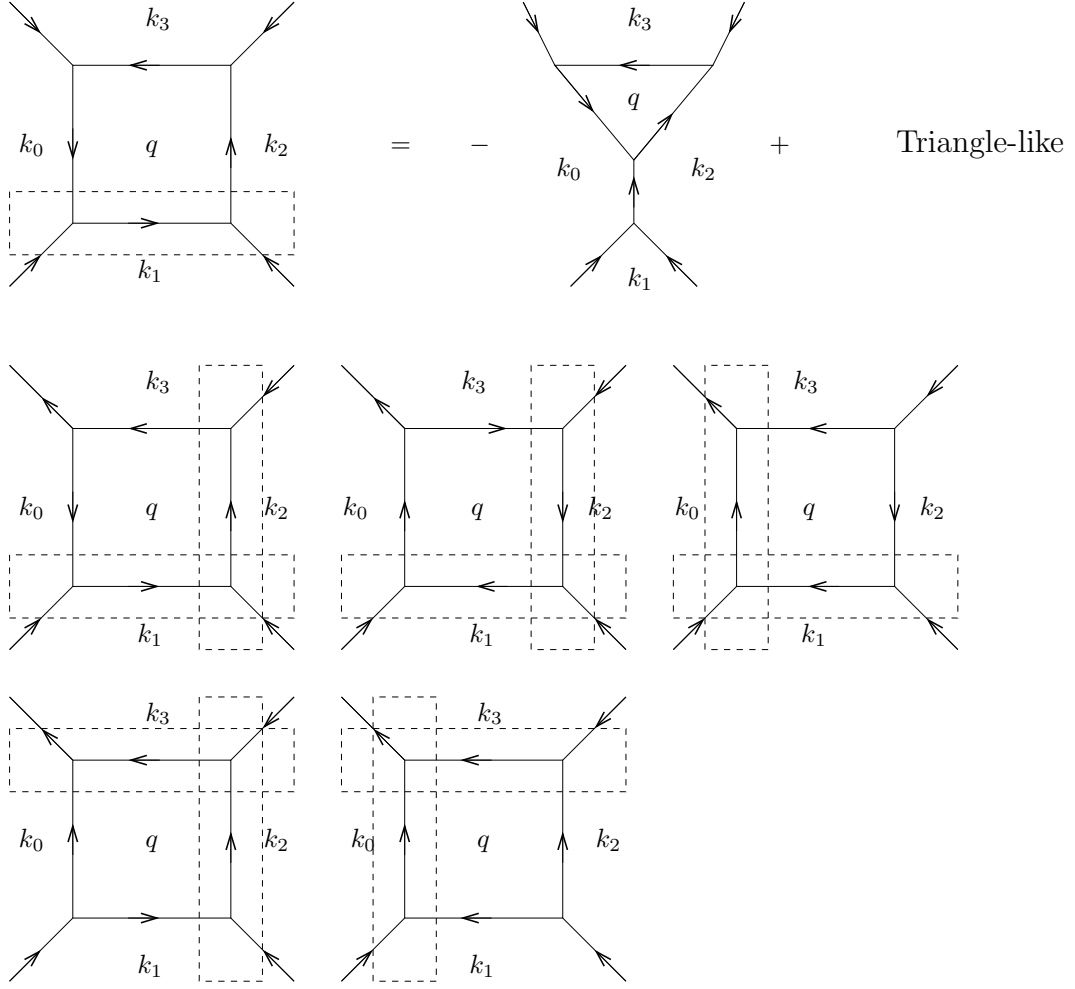}
\caption{The boxes for the finite one-loop amplitudes can be reduced to
triangles and triangle-like integrals by replacing any of the sub-diagrams
enclosed in a dashed box. A typical replacement is shown in the first line.}
\label{boxreduction1}
\end{center}
\end{figure}

We have seen that the on-shell limit of a tree amplitude is dramatically
simpler than the off-shell expression. We can identify 
tree amplitudes as sub-diagrams of one loop diagrams,
but some of the legs of these sub-diagrams will be off-shell, so it
would seem that the simplifying features of the on-shell limit
can't be exploited. However, if one leaves the denominators of the
trees in their original covariant form, then the numerators can always be
written as the simplified on-shell expression plus terms each of
which contain at least one factor of the virtuality $p^2$ of one
of the off-shell legs. In a box diagram such terms will
cancel a propagator reducing
the required loop integrand to one with the structure of a
triangle diagram. Since triangle integrals are considerably easier to
analyze than box integrals, the resulting simplification is very useful.
In this section, we use this technique to reduce the complications
of the box diagrams for the helicity configurations $\wedge\wedge\wedge\wedge$
and $\wedge\wedge\wedge\vee$, which are the focus of this article.
For these spin configurations we can always find a four point
sub-diagram of the box diagram that involves helicities
$\wedge\wedge\wedge\vee$ or $\wedge\vee\vee\vee$, as indicated
in Figure \ref{boxreduction1}. Thus these box diagrams can be completely
reduced to triangle-type integrals.
\begin{subsection}
{\large $\wedge\wedge\wedge\wedge$}
\end{subsection}
Referring to the first line of Fig.~\ref{boxreduction1}, and remembering
that there is one other box contribution with the arrows
circulating clockwise with the same value, we read off the
box contribution to the four like helicity process:
\bea
\Gamma_{\rm Box}^{\wedge\wedge\wedge\wedge}
&=&2(2g)^4N_c^2
\int {d^4q\over16\pi^4}{K_{12}^\wedge K_{35}^\wedge K_{64}^\wedge
\over p_1^+p_2^+p_3^+ p_4^+p_{12}^2}
\left[{K_{25}^\wedge\over q_1^2q_2^2q_3^2}+{K_{61}^\wedge\over q_0^2q_1^2q_3^2}
-{K_{61}^\wedge+K_{25}^\wedge-K_{12}^\wedge\over q_0^2q_2^2q_3^2}\right]\label{uuuu}
\eea
where the off-shell momenta are defined as $p_5=q-k_2, p_6=k_0-q$ and  $q_i^2=(q-k_i)^2$.
\begin{subsection}
{\large $\wedge\wedge\wedge\vee$}
\end{subsection}
Here we gather the triangle-like contributions arising from the
five box diagrams contributing
to the process with three like helicities (see 
the second and third lines of Fig.~\ref{boxreduction1}). 
In the loop integrand, there are four distinct 
triangle denominator structures
descended from the box denominators $(q_0^2q_1^2q_2^2q_3^2)^{-1}$.
We list the coefficient of each structure.
The coefficient of $g^4N_c^2(\pi^4q_0^2q_1^2q_3^2)^{-1}$ is
\bea
&&{K_{12}^{\wedge}K_{61}^\wedge K_{64}^\vee
K_{35}^{\wedge}\over p_1^+p_2^+p_3^+p_4^+p_{12}^2}
\left[{q^{+2}\over (q^++p_{4}^+)^2}+{(q^++p_{4}^+)^2\over q^{+2}}\right]
+{K_{12}^{\wedge}K_{61}^\wedge K_{64}^\wedge
K_{35}^{\vee}\over p_1^+p_2^+p_3^+p_4^+p_{12}^2}{p_{3}^{+2}p_{4}^{+2}
\over(q^+-p_{12}^+)^2(q^++p_{4}^+)^2} 
\nonumber\\
&&\qquad+{K_{34}^{\wedge}K_{61}^\wedge K_{64}^\wedge
K_{25}^{\vee}\over p_1^+p_2^+p_3^+p_4^+p_{12}^2}{p_{2}^{+2}p_{4}^{+2}
\over(q^+-p_1^+)^2(q^+-p_{12}^+)^2}+{K_{34}^{\wedge}K_{61}^\vee K_{64}^\wedge
K_{25}^{\wedge}\over p_1^+p_2^+p_3^+p_4^+p_{12}^2}{p_{1}^{+2}p_{4}^{+2}
\over(q^+-p_1^+)^2q^{+2}}.\label{T013}
\eea
The coefficient of $g^4N_c^2(\p^4q_1^2q_2^2q_3^2)^{-1}$ is
\bea
&&{K_{12}^{\wedge}K_{25}^\wedge K_{64}^\vee
K_{35}^{\wedge}\over p_1^+p_2^+p_3^+p_4^+p_{12}^2}
\left[{q^{+2}\over (q^++p_{4}^+)^2}+{(q^++p_{4}^+)^2\over q^{+2}}\right]
+{K_{12}^{\wedge}K_{25}^\wedge K_{64}^\wedge
K_{35}^{\vee}\over p_1^+p_2^+p_3^+p_4^+p_{12}^2}{p_{3}^{+2}p_{4}^{+2}
\over(q^+-p_{12}^+)^2(q^++p_{4}^+)^2}
\nonumber\\
&&\qquad+{K_{34}^{\wedge}K_{35}^\wedge K_{61}^\wedge
K_{25}^{\vee}\over p_1^+p_2^+p_3^+p_4^+p_{12}^2}{p_{2}^{+2}p_{4}^{+2}
\over(q^+-p_1^+)^2(q^+-p_{12}^+)^2}+{K_{34}^{\wedge}K_{61}^\vee K_{35}^\wedge
K_{25}^{\wedge}\over p_1^+p_2^+p_3^+p_4^+p_{12}^2}{p_{1}^{+2}p_{4}^{+2}
\over(q^+-p_1^+)^2q^{+2}}.\label{T123}
\eea
The coefficient of $g^4N_c^2(\pi^4q_0^2q_1^2q_2^2)^{-1}$ is
\bea
-{K_{34}^{\wedge}K_{61}^\wedge K_{65}^\wedge
K_{25}^{\vee}\over p_1^+p_2^+p_3^+p_4^+p_{12}^2}{p_{2}^{+2}p_{4}^{+2}
\over(q^+-p_1^+)^2(q^+-p_{12}^+)^2}-{K_{34}^{\wedge}K_{61}^\vee K_{65}^\wedge
K_{25}^{\wedge}\over p_1^+p_2^+p_3^+p_4^+p_{12}^2}{p_{1}^{+2}p_{4}^{+2}
\over(q^+-p_1^+)^2q^{+2}}.\label{T012}
\eea
The coefficient of $g^4N_c^2(\pi^4q_0^2q_2^2q_3^2)^{-1}$ is
\bea
-{K_{12}^{\wedge}K_{56}^\wedge K_{64}^\vee
K_{35}^{\wedge}\over p_1^+p_2^+p_3^+p_4^+p_{12}^2}
\left[{q^{+2}\over (q^++p_{4}^+)^2}+{(q^++p_{4}^+)^2\over q^{+2}}\right]
-{K_{12}^{\wedge}K_{56}^\wedge K_{64}^\wedge
K_{35}^{\vee}\over p_1^+p_2^+p_3^+p_4^+p_{12}^2}{p_{3}^{+2}p_{4}^{+2}
\over(q^+-p_{12}^+)^2(q^++p_{4}^+)^2}.\label{T023}
\eea

\section{$\wedge\wedge\wedge\wedge$ at One Loop}
\begin{subsection}
{Direct Calculation of the Box and Triangle contributions}
\end{subsection}
Because the tree contribution to this process is
zero, the one loop contribution must be completely finite.
The triangle-like integral  descended from the box diagrams is spelled out in
Eq. (\ref{uuuu}). The integral over q is finite and can be explicitly 
evaluated. The last term in Eq. (\ref{uuuu}) can be identified 
(note that $K_{61}^\wedge+K_{25}^\wedge-K_{12}^\wedge=K_{56}^\wedge$)
as the negative of the
pure triangle diagram attached to legs 3 and 4. So, when added to 
the triangle diagrams 
attached to legs 3 and 4, this term is canceled out  
so the box and this triangle diagram together become
\be
 g^4N_c^2\int {d^4q\over \pi^4}{2 K_{12}^\wedge K_{64}^\wedge K_{35}^\wedge \over
p_1^+p_2^+ p_3^+p_4^+p_{12}^2 } \left[
{K_{61}^\wedge \over q_0^2 q_1^2 q_3^2 }+{ K_{25}^\wedge \over q_1^2q_2^2 q_3^2}\right], 
\ee
which, after integration over $q$, gives.
\be
{g^4N_c^2\over 3 \pi^2}{K_{12}^\wedge\over 
p_1^+p_2^+ p_3^+p_4^+p_{12}^2 p_{14}^2}
\big [ K_{41}^\wedge K_{23}^\wedge (K_{41}^\wedge +K_{23}^\wedge)+
K_{34}^\wedge ({K_{41}^\wedge}^2 +{K_{23}^\wedge}^2) \big ].
\ee
The remaining triangles are also finite and can easily be calculated.
The triangle attached to legs 1 and 2 is
\be
{g^4N_c^2\over 3 \pi^2}{K_{34}^\wedge {K_{12}^\wedge}^3 \over 
p_1^+p_2^+ p_3^+p_4^+p_{12}^4}.
\ee
The triangle attached to legs 1 and 4 is
\be
{g^4N_c^2\over 3 \pi^2}{K_{23}^\wedge {K_{41}^\wedge}^3 \over 
p_1^+p_2^+ p_3^+p_4^+p_{14}^4}.
\ee
And the triangle attached to legs 2 and 3 is
\be
{g^4N_c^2\over 3 \pi^2}{K_{41}^\wedge {K_{23}^\wedge}^3 \over 
p_1^+p_2^+ p_3^+p_4^+p_{14}^4}.
\ee
As explained in the next subsection,
the physical one loop scattering amplitude contains no self energy
insertions, and so is obtained by  
adding all these contributions. So, the {\it physical} one loop 
scattering amplitude is
\bq
\Gamma^{\wedge\wedge\wedge\wedge}&=&
{g^4N_c^2\over 3 \pi^2}{1\over p_1^+p_2^+ p_3^+p_4^+}\bigg[
{K_{12}^\wedge K_{41}^\wedge K_{23}^\wedge (K_{41}^\wedge +K_{23}^\wedge)+
K_{12}^\wedge K_{34}^\wedge ({K_{41}^\wedge}^2 +{K_{23}^\wedge}^2) 
\over p_{12}^2 p_{14}^2}\nonumber \\
&~&~~~~~~~~~~~~~~~~~~~~+{K_{34}^\wedge {K_{12}^\wedge}^3 \over p_{12}^4}
+{K_{23}^\wedge {K_{41}^\wedge}^3 \over p_{14}^4} +
{K_{41}^\wedge {K_{23}^\wedge}^3 \over p_{14}^4 }
 \bigg ]\nonumber \\
&=& {g^4N_c^2\over 3 \pi^2}
{1\over p_1^+p_2^+ p_3^+p_4^+ p_{12}^2  p_{14}^2}\big[
K_{12}^\wedge K_{41}^\wedge K_{23}^\wedge (K_{41}^\wedge +K_{23}^\wedge)+
K_{12}^\wedge K_{34}^\wedge ({K_{41}^\wedge}^2 
+{K_{23}^\wedge}^2) \nonumber \\
&~&~~~~~~~~~~~~~~~~~~~~-{K_{12}^\wedge}^2 K_{23}^\wedge K_{41}^\wedge
-{K_{41}^\wedge}^2 K_{12}^\wedge K_{34}^\wedge -{K_{23}^\wedge}^2 K_{12}^\wedge
K_{34}^\wedge  \big] \nonumber \\
&=& {g^4N_c^2\over 3 \pi^2}{K_{12}^\wedge K_{41}^\wedge K_{23}^\wedge \over
p_1^+p_2^+ p_3^+p_4^+p_{12}^2  p_{14}^2 } 
(K_{41}^\wedge+K_{23}^\wedge-K_{12}^\wedge)\nonumber \\
&=&{g^4\over 3 \pi^2} {K_{12}^\wedge K_{23}^\wedge K_{34}^\wedge K_{41}^\wedge
\over p_1^+p_2^+ p_3^+p_4^+ p_{12}^2  p_{14}^2 }\nonumber \\
&=&-{g_s^4N_c^2 \over 48 \pi^2}{K_{21}^\wedge K_{43}^\wedge 
\over K_{21}^\vee K_{43}^\vee}
\label{4likeamp}
\eq
where $g_s=g\sqrt2$ is the conventional QCD coupling constant. 
Removing a factor of $N_c$, we find that this result 
agrees exactly with the known one \cite{bernk,kunsztst}.

\begin{subsection}
{Discussion and a Remarkable Identity}
\end{subsection}
In the previous subsection we identified the physical scattering 
amplitude with the box and triangle diagrams only. However, 
our $\delta$ regulator
leads to a non-zero result for the one loop helicity flipping self energy 
function $\Pi^{\wedge\wedge}$:
\bea
\Pi^{\wedge\wedge}(k,k^\prime)&=&{g^2N_c\over12\pi^2}(k^{\wedge2}
+k^{\wedge}k^{\prime\wedge}+k^{\prime\wedge2})
\eea
Here $k,k^\prime$ are the dual momenta assigned to the two external regions
separated by the external lines.
(With our definitions the conventional $\alpha_s=g^2/2\pi$,
or in other words $g=g_s/\sqrt2$.)
This result for $\Pi^{\wedge\wedge}$  is anomalous in several
respects. 
For one thing it violates the translational
symmetry $k,k^\prime\to k+a,k^\prime+a$.
This formal symmetry of the unregulated theory is violated by
our $\delta$ regulator. It does not disappear as $\delta\to0$
because the self energy integral is quadratically divergent
by power-counting, so terms of order $O(\delta)$ get multiplied
by a factor $1/\delta$ and so survive the limit. These regulator
artifacts are not present in the triangle and box diagrams
for this process.
In addition, the fact that $\Pi^{\wedge\wedge}\neq0$ 
on shell would imply a non-zero transition
amplitude for a gluon's helicity to flip, which is inconsistent
with Lorentz covariance. Fortunately, the result is a finite 
quadratic polynomial in $k,k^\prime$, which can be canceled
by a local counterterm. To achieve Lorentz invariance this counterterm
must be tuned so that $\Pi_{\rm TOT}^{\wedge\wedge}=0$, and this justifies
the identification of the physical scattering amplitude
we made in the previous subsection.

\begin{figure}[ht]
\psfrag{'q'}{$q$}
\psfrag{'k0'}{$k_0$}
\psfrag{'k1'}{$k_1$}
\psfrag{'k2'}{$k_2$}
\psfrag{'k3'}{$k_3$}
\begin{center}
\includegraphics[width=1.5in]{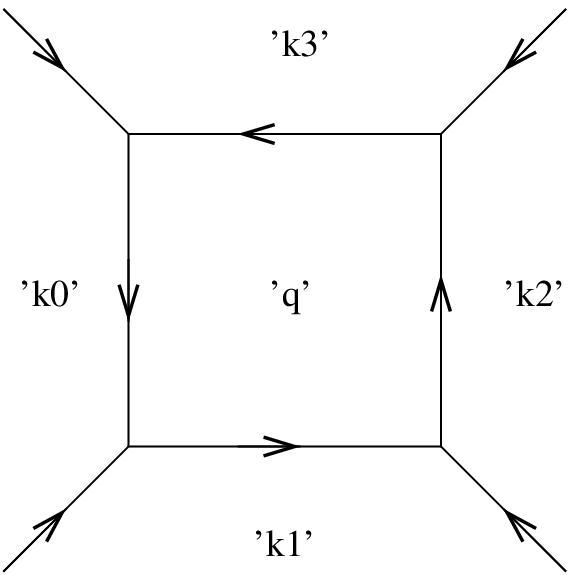}
\hskip1in\includegraphics[width=3in]{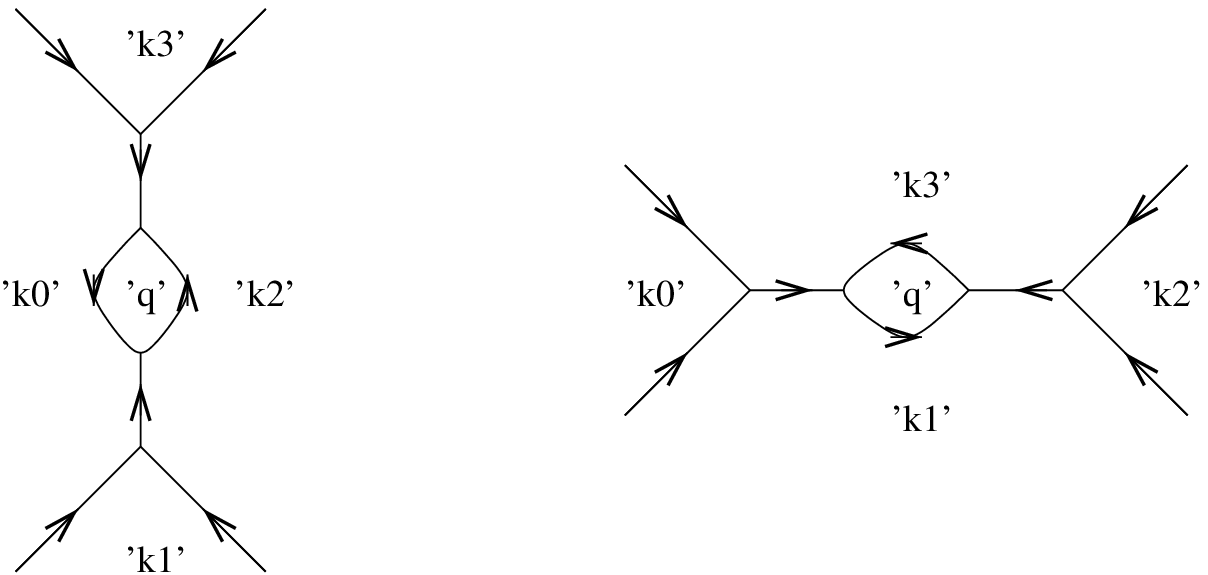}\\
\includegraphics[width=6in]{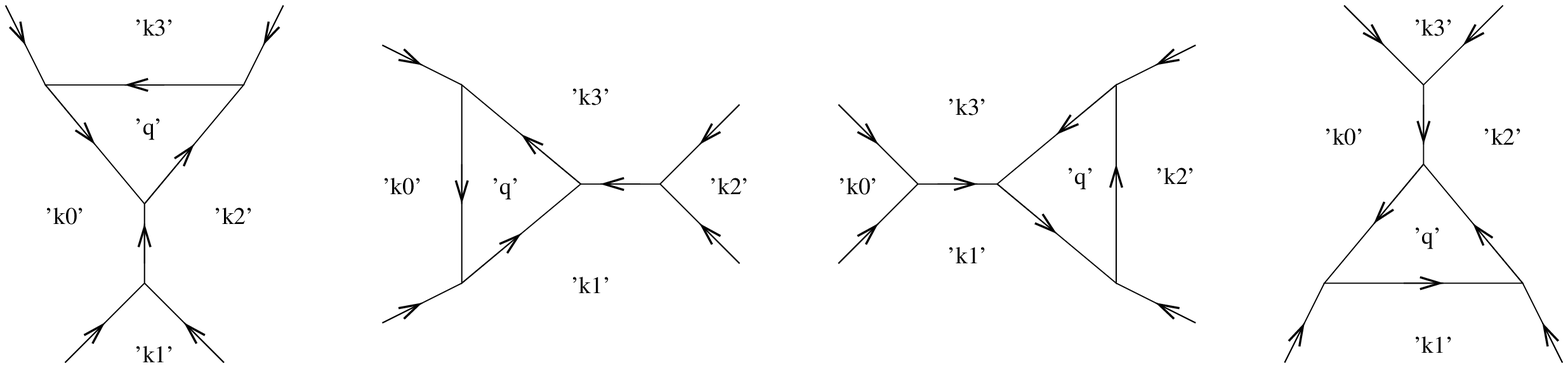}\\
\caption{The diagrams contributing to the four point like helicity
Green function without bubbles on external legs.}
\label{4ptgreen1}
\end{center}
\end{figure}

\begin{figure}[ht]
\psfrag{'q'}{$q$}
\psfrag{'k0'}{$k_0$}
\psfrag{'k1'}{$k_1$}
\psfrag{'k2'}{$k_2$}
\psfrag{'k3'}{$k_3$}
\begin{center}
\includegraphics[width=6in]{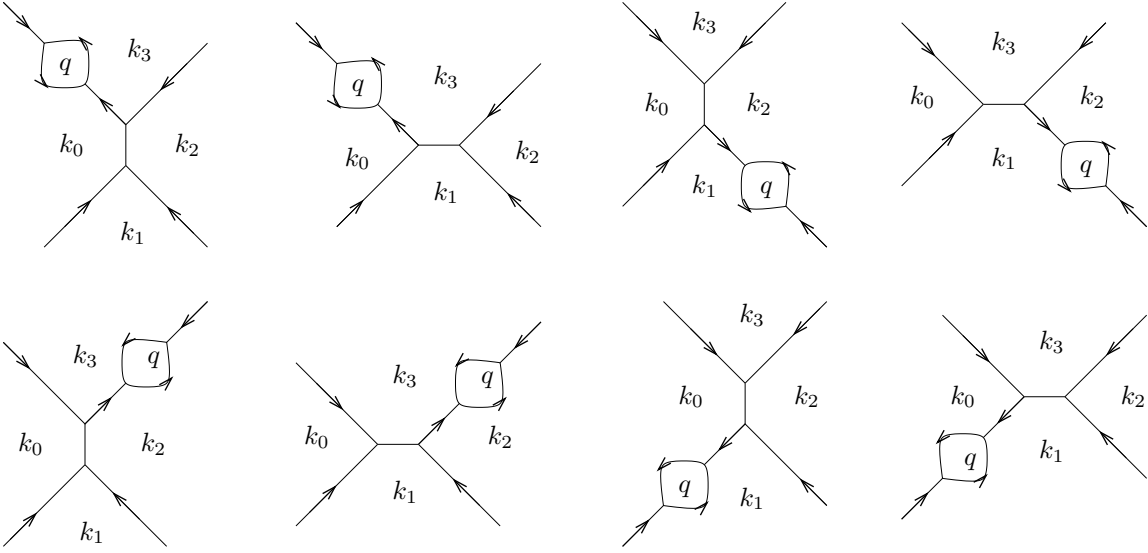}
\caption{The diagrams contributing to the four point like helicity
Green function with bubbles on external legs.}
\label{4ptgreen2}
\end{center}
\end{figure}

If we don't include this counterterm, however, 
there is a remarkable property of the one loop integrand
for the complete Green function corresponding
to this all like-helicity amplitude in the on-shell limit:
It is identically zero \cite{bernprivate}!
To be precise, this means that the sum of the integrands of
the box, four triangles, two self-energy bubbles
on internal lines and eight self energy bubbles on external lines
(see Figs.~\ref{4ptgreen1},\ref{4ptgreen2}), {\it with a particular routing
of momenta through the individual diagrams}, is identically zero
if all external legs are put on-shell.
It is crucial here to include the diagrams with self-energy insertions on
external legs, which have a finite
on-shell limit because the pole due to the internal line
attached to the bubble is canceled by a zero in the on-shell
tree amplitude with three like helicities. Because of this remarkable
identity, we can interpret the all like helicity scattering amplitude
as a pure anomaly arising from the need for counterterms that
restore Lorentz invariance. Without the counterterms, the amplitude
would vanish, but Lorentz invariance would be violated
for other physical processes.\footnote{In
covariant gauge calculations using dimensional regularization, this
fact is understood  in the following way: The loop integrand vanishes only
in four space-time dimensions and only like a single power of
$D-4$. UV divergences in the loop integrals become simple
poles at $D=4$. These poles are canceled by the zero of the
integrand at $D=4$, rendering the amplitude finite but non-zero.}

The routing of loop momenta in the individual diagrams that ensures the
vanishing of the integrand is indeed the routing dictated by the
worldsheet representation \cite{bardakcit}. Namely, each diagram
divides the plane into the same number of regions, and the dual momentum
for each region is assigned identically for each diagram. Then the
regulator factor $e^{-\delta{\bfs q}^2}$ is the same for each diagram
and doesn't disturb the complete cancellation of the integrand.

To confirm that the sum of the integrands vanishes
with the indicated routing of momenta, we note that this sum
is a meromorphic function of $q^-$ that vanishes at infinity.
It has poles where the denominators vanish with residues that 
are just on-shell 6 point trees with the pair of legs that
correspond to the internal line with the pole under consideration 
carrying equal and opposite momenta. Since the set of diagrams
included does not include tadpole insertions, the six point diagram
where these two legs share a common cubic vertex is absent.
But these diagrams are zero because of this momentum 
constraint\footnote{To see this first evaluate the off-shell 
($p_5^2\neq0$) five-point function,
which is easily found to be
\bea
A_{\wedge\wedge\wedge\wedge\vee}^{tree}=-{p_2^+p_3^+p_5^+\over
8K_{43}^\vee K_{32}^\vee K_{21}^\vee}p_5^2
\eea
which of course vanishes on shell. We then get these diagrams by
multiplying by $-K_{76}^\wedge p_7^+/(p_5^+p_6^+p_5^2)$, which is the
$567$ vertex times the propagator for leg $5$. Thus residue of the
pole in the tadpole integrand is the $p_7\to-p_6$ limit
of
\bea
R_{\rm tadpole}={p_2^+p_3^+p_7^+\over
8p_6^+ K_{43}^\vee K_{32}^\vee K_{21}^\vee}K_{76}^\wedge\to0
\eea}.
Since five legs of the six point trees giving the
one loop residues have the same helicity,
they all vanish on shell, so all the poles of the meromorphic function have 
zero residue. This implies that the function is identically
zero, and we have seen that residues of the tadpole integrands vanish by 
themselves so they can be deleted.

We close this section by demonstrating these results by
direct calculation. The four poles in $q^-$ are given
by $(q-k_i)^2=0$ for each $i=0,1,2,3$. If the residue of any
one of these poles is zero, it follows from symmetry that the
other three residues are also zero. So we need only consider the
first pole with $i=0$. The diagrams that give this pole are the box,
three of the triangles, one of the internal self energies and four of the
external self energies: namely
all the diagrams with an internal line bordering
the region labeled $k_0$. The cyclic ordering of the tree is then
$123456$ where $1234$ label the external legs of the loop diagram and
$p_5=-p_6=q-k_0$. 

Using our expression for the off-shell four point function it is
convenient to group the contributing diagrams into: 
\begin{enumerate}
\item The four
external self energy diagrams which sum and  simplify to
\bea
{p_5^{+2}\over8}\left[-{p_3^+\over p_5^+}{K_{51}^\wedge\over
K_{51}^\vee K_{43}^\vee K_{32}^\vee}+{p_2^+\over p_5^+}{K_{54}^\wedge\over
K_{54}^\vee K_{21}^\vee K_{32}^\vee}\right]
\eea
where we have made use of $p_6=-p_5$. Notice that the explicit
factor of $p^2$ in the off-shell four point function cancels the
propagator factor $1/p^2$ so this limit is finite. Here 
$p=p_i+p_5+p_6$ if the bubble is on the $i$th leg.
\item The four diagrams that also have the propagator $(q-k_2)^{-2}$
(the box, two of the triangles and the internal self energy). 
These diagrams all together give the product of two off-shell
four point functions times this propagator:
\bea
{p_5^{+2}\over8}\left[{p_1^+p_4^+(p_3+p_4+p_5)^2\over
2K_{51}^\vee K_{21}^\vee K_{43}^\vee K_{45}^\vee}\right]
\eea
\item The third triangle diagram:
\bea
{p_5^{+2}\over8}\left[{K_{21}^\wedge+K_{52}^\wedge+K_{31}^\wedge
+K_{53}^\wedge \over
K_{51}^\vee K_{32}^\vee K_{54}^\vee }\right]
\eea
\end{enumerate}
We have omitted coupling constant factors $(2g\sqrt{N_c})^4$ in the above
expressions.
The claim is that the sum of all these contributions is zero.
Putting them all over a common denominator, we see immediately that
the terms independent of $p_5$ sum to zero:
\bea
(K_{21}^\wedge+K_{31}^\wedge)K_{21}^\vee K_{43}^\vee
-{p_1^+p_4^+(p_3+p_4)^2K_{32}^\vee\over2}
=K_{43}^\vee p_1^+\left(-{1\over p_1^+}K_{41}^\wedge K_{21}^\vee
+{1\over p_3^+}K_{43}^\wedge
K_{32}^\vee\right)=0
\eea
This leaves 
\bea
{p_5^{+2}\over8K_{51}^\vee K_{21}^\vee K_{43}^\vee K_{32}^\vee K_{54}^\vee}
\bigg[-{p_3^+\over p_5^+}(K_{51}^\wedge K_{21}^\vee K_{54}^\vee)
+{p_2^+\over p_5^+}(K_{54}^\wedge K_{51}^\vee K_{43}^\vee)\nonumber\\
-p_1^+p_4^+p_5\cdot(p_3+p_4)K_{32}^\vee
+(K_{52}^\wedge+K_{53}^\wedge)K_{21}^\vee K_{43}^\vee\bigg]
\eea
A systematic way to see that these terms also cancel is to
use the identities
\bea
p_4^+K_{51}^\vee+p_1^+K_{45}^\vee+p_5^+K_{14}^\vee&=&0\\
p_4^+K_{53}^\vee+p_3^+K_{45}^\vee+p_5^+K_{34}^\vee&=&0
\eea
to isolate the dependence on $p_5^\vee$. Then the quantity in square 
brackets reads
\bea
\left[\phantom{\int}\right]
=K_{54}^\vee{p_1^+p_3^+\over p_4^+}\sum_i {K_{5i}^\wedge K_{i2}^\wedge
\over p_i^+}+K_{43}^\vee\left[p_2^+\sum_i {K_{5i}^\wedge K_{i1}^\wedge
\over p_i^+}+{K_{53}^\wedge\over p_3^+}\left(p_3^+K_{21}^\vee+
p_1^+K_{32}^\vee+p_2^+K_{13}^\vee\right)\right]=0
\eea
As already stated, symmetry dictates that the residues of the other
three poles in $q^-$ also vanish, which then proves that the
integrand vanishes identically at all $q$. If we regulate the
loop integrals by a direct cutoff on $q$, for example by inserting
a factor $e^{-\delta {\bfs q}^2}$, it follows that the
sum of all the regulated integrals vanishes also.

Incorporating this counterterm, the revised on shell MHV Green function,
namely value of the box and four triangle contributions combined,
can be identified with the physical scattering amplitude.
In view of the fact that the unsubtracted diagrams sum to zero,
the physical scattering amplitude is just the negative of the
self energy contributions to the on-shell Green function.
Thus we can calculate the scattering amplitude by a
purely algebraic combination of the numerous self-energy insertions.
For the reader's entertainment
We list the evaluation of these simple diagrams.
\begin{enumerate}
\item Bubbles on internal lines: 
\bea
{(p_1^++p_2^+)^2K_{14}^\wedge K_{32}^\wedge(k_0^{\wedge2}
+k_0^{\wedge}k_2^{\wedge}+k_2^{\wedge2})+(p_1^++p_4^+)^2
K_{21}^\wedge K_{43}^\wedge(k_1^{\wedge2}
+k_1^{\wedge}k_3^{\wedge}+k_3^{\wedge2})\over p_1^+p_2^+p_3^+p_4^+st}
\eea
\item Bubble on leg 4:
\bea
{-p_4^{+2}K_{32}^\wedge
K_{21}^\wedge(k_0^{\wedge2}
+k_0^{\wedge}k_3^{\wedge}+k_3^{\wedge2})\over p_1^+p_2^+p_3^+p_4^+st}
={[-p_4^+p_3^+K_{14}^\wedge K_{21}^\wedge+p_4^+(p_3^++p_2^+)K_{43}^\wedge
K_{21}^\wedge](k_0^{\wedge2}
+k_0^{\wedge}k_3^{\wedge}+k_3^{\wedge2})\over p_1^+p_2^+p_3^+p_4^+st}
\eea
\item Bubble on leg 3:
\bea
{-p_3^{+2}K_{14}^\wedge
K_{21}^\wedge(k_2^{\wedge2}
+k_2^\wedge k_3^\wedge+k_3^{\wedge2})
\over p_1^+p_2^+p_3^+p_4^+st}
\eea
\item Bubble on leg 2:
\bea
{-p_2^{+2}K_{14}^\wedge
K_{43}^\wedge(k_1^{\wedge2}
+k_1^{\wedge}k_2^{\wedge}+k_2^{\wedge2})\over p_1^+p_2^+p_3^+p_4^+st}
={[-p_2^+p_3^+K_{14}^\wedge K_{21}^\wedge-p_2^+(p_1^++p_2^+)K_{14}^\wedge
K_{32}^\wedge](k_1^{\wedge2}
+k_1^{\wedge}k_2^{\wedge}+k_2^{\wedge2})\over p_1^+p_2^+p_3^+p_4^+st}
\eea
\item Bubble on leg 1:
\bea
&&{-p_1^{+2}K_{43}^\wedge
K_{32}^\wedge(k_0^{\wedge2}
+k_0^{\wedge}k_1^{\wedge}+k_1^{\wedge2})\over p_1^+p_2^+p_3^+p_4^+st}
=\nonumber\\
&&\qquad{[-p_1^+p_3^+K_{14}^\wedge K_{21}^\wedge-p_1^
+(p_1^++p_4^+)K_{43}^\wedge
K_{21}^\wedge-p_1^+(p_1^++p_2^+)K_{14}^\wedge K_{32}^\wedge](
k_0^{\wedge2}
+k_0^{\wedge}k_1^{\wedge}+k_1^{\wedge2})\over p_1^+p_2^+p_3^+p_4^+st}
\eea
\end{enumerate}
We simplify the numerator in sum of all these diagrams in stages
\begin{enumerate}
\item Coefficient of $K_{14}^\wedge K_{32}^\wedge$:
\bea
(p_1^++p_2^+)(k_0^\wedge+k_1^\wedge+k_2^\wedge)(p_1^+(k_2^\wedge-k_1^\wedge)
+p_2^+(k_0^\wedge-k_1^\wedge))=-K_{21}^\wedge(p_1^++p_2^+)(k_0^\wedge
+k_1^\wedge+k_2^\wedge)
\eea
\item Coefficient of $K_{43}^\wedge K_{21}^\wedge$:
\bea
(p_1^++p_4^+)(k_0^\wedge+k_1^\wedge+k_3^\wedge)(p_1^+(k_3^\wedge-k_0^\wedge)
+p_4^+(k_1^\wedge-k_0^\wedge))=-K_{14}^\wedge(p_1^++p_4^+)(k_0^\wedge
+k_1^\wedge+k_3^\wedge)
\eea
\item Coefficient of $K_{14}^\wedge K_{21}^\wedge$:
\bea
&&\hskip-.5in
-p_3^+[p_4^+(k_0^{\wedge2}+k_0^{\wedge}k_3^{\wedge}+k_3^{\wedge2})
+p_3^+(k_2^{\wedge2}+k_2^{\wedge}k_3^{\wedge}+k_3^{\wedge2})
+p_2^+(k_1^{\wedge2}+k_1^{\wedge}k_2^{\wedge}+k_2^{\wedge2})
+p_1^+(k_0^{\wedge2}+k_0^{\wedge}k_1^{\wedge}+k_1^{\wedge2})]\nonumber\\
&&\hskip-.5in
=-p_3^+[K_{43}^\wedge(k_0^{\wedge}+k_2^{\wedge}+k_3^{\wedge})
+K_{21}^\wedge(k_0^\wedge+k_1^{\wedge}+k_2^{\wedge})]
\eea
\end{enumerate}
Putting the numerator all together we have
\bea
N&=&K_{14}^\wedge K_{21}^\wedge[-K_{32}^\wedge(p_1^++p_2^+)(k_0^\wedge
+k_1^\wedge+k_2^\wedge)-K_{43}^\wedge(p_1^++p_4^+)(k_0^\wedge
+k_1^\wedge+k_3^\wedge)\nonumber\\
&&-p_3^+(K_{43}^\wedge(k_0^{\wedge}+k_2^{\wedge}+k_3^{\wedge})
+K_{21}^\wedge(k_0^\wedge+k_1^{\wedge}+k_2^{\wedge}))
]\nonumber\\
&=&K_{14}^\wedge K_{21}^\wedge K_{43}^\wedge 
[p_2^+(k_3^\wedge-k_2^\wedge)+p_3^+(k_1^\wedge-k_2^\wedge)]
=-K_{14}^\wedge K_{21}^\wedge K_{43}^\wedge K_{32}^\wedge 
\eea
So, restoring the coupling factors,
 the physical scattering amplitude is the negative of this
\bea
\Gamma^{\wedge\wedge\wedge\wedge}&=&
+N_c(2g)^2{g^2N_c\over12\pi^2}
{K_{14}^\wedge K_{21}^\wedge K_{43}^\wedge K_{32}^\wedge
\over p_1^+p_2^+p_3^+p_4^+st}=-N_c(2g)^2{g^2N_c\over12\pi^2}
{K_{21}^{\wedge2} K_{43}^{\wedge2}
\over p_1^+p_2^+p_3^+p_4^+s^2}\nonumber\\
&=&-{g^4N_c^2\over12\pi^2}
{K_{21}^{\wedge} K_{43}^{\wedge}
\over K_{21}^\vee K_{43}^\vee}=-{g_s^4N_c^2\over48\pi^2}
{K_{21}^{\wedge} K_{43}^{\wedge}
\over K_{21}^\vee K_{43}^\vee}
\eea
in agreement with (\ref{4likeamp}).
\section{$\wedge\wedge\wedge\vee$ at One Loop}
\begin{subsection}
{Cubic Vertex corrections and self energy insertions on internal
lines.}
\end{subsection}
The one loop corrections to the cubic vertices 
have the topology of triangle and swordfish
diagrams. They have been evaluated in
\cite{thornlcnotes} and quoted in Section 6.   
With the definitions given in Section 6 we now summarize the cubic vertex
corrections to on-shell scattering of glue by glue:
\bea
{\Gamma^{\wedge\wedge\wedge\vee}_\triangle\over16g^4N_c^2}
&=& -{p_4^+\over32\pi^2p_1^+p_2^+p_3^+}\bigg\{
{K_{34}^\wedge K_{12}^\wedge\over p_{12}^2}
\left[{22\over3}\ln(p_{12}^2e^\gamma\delta)-{140\over9}
-S_3^{q^+}(p_1,p_2)-S_3^{q^+}(-p_4,-p_3)
+{p_1^+p_2^+\over 3p_{12}^{+2}}\right]\nonumber\\
&&+{K_{23}^\wedge K_{41}^\wedge\over p_{14}^2}
\left[{22\over3}\ln(p_{14}^2e^\gamma\delta)-{140\over9}
-S_2^{q^+}(-p_4,-p_{23})-S_1^{q^++p_4^+}(p_{14},p_2)
+{p_2^+p_3^+\over 3p_{14}^{+2}}\right]\bigg\}
\nonumber\\
&&+{p_3^+\over48\pi^2p_1^+p_2^+p_4^+p_{12}^{+2}}
{K_{12}^{\wedge3}K_{34}^\vee\over p_{12}^4}
+{p_1^+\over48\pi^2p_2^+p_3^+p_4^+p_{14}^{+2}}
{K_{23}^{\wedge3}K_{41}^\vee\over p_{14}^4}
\eea
where we stress that the cubic counterterm has been included.
Finally we give the self energy insertions on internal lines obtained
in Section 5,
again including the counterterm that sets to zero the
helicity flip self energy as well as the boundary counterterm and
the gluon mass counterterm:
\bea
{\Gamma^{\wedge\wedge\wedge\vee}_{\rm SE}\over16g^4N_c^2}
&=&-{p_4^+\over16\pi^2p_1^+p_2^+p_3^+}\bigg\{
{K_{34}^\wedge K_{12}^\wedge\over p_{12}^2}
\left[{\cal A}^{q^+}(p_{12}^2,p_{12}^+)-{11\over6}\ln(p_{12}^2e^\gamma\delta)  
+{67\over18}\right]\nonumber\\
&&\hskip1in +{K_{23}^\wedge K_{41}^\wedge\over p_{14}^2}
\left[{\cal A}^{q^++p_4^+}(p_{14}^2,p_{14}^+)
-{11\over6}\ln(p_{14}^2e^\gamma\delta)
+{67\over18}\right]\bigg\}\\
{\cal A}^{q^+}(p^2,p^+)&=&\sum_{q^+}\left[{1\over q^+}+{1\over p^+-q^+}\right]
\ln\left\{{q^+\over p^+}\left(1-{q^+\over p^+}\right)p^2e^\gamma\delta
\right\}
\eea
where the superscript on ${\cal A}$ signifies that longitudinal momentum
on the internal line on the left with $p^+>0$ and time running
up.

\begin{figure}[ht]
\begin{center}
\includegraphics[width=11cm]{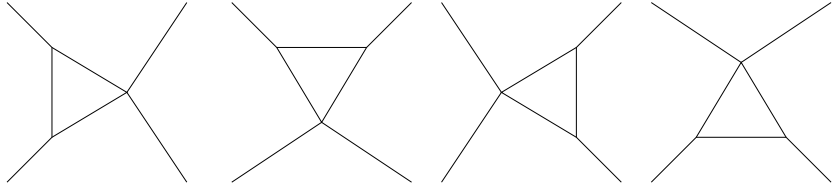}
\caption{The quartic triangle diagrams shown generically, 
without arrows indicating
spin flow. Particle labels $1234$ are applied counter-clockwise
starting at the lower left of each diagram.}
\end{center}
\label{quarttri}
\end{figure}

\begin{subsection}
{Quartic Triangle Diagrams}
\end{subsection}
The four topologies of ``quartic triangle'' diagrams are shown in
Fig.~\ref{quarttri}. We sketch their evaluation here with more details
found in \cite{thornlcnotes}. 
The coefficient of $16g^4N_c^2(16\pi^4q_0^2q_1^2q_3^2)^{-1}$ in the
loop integrand is
\bea
K_{61}^\wedge K_{64}^\wedge
{p_4^+[p_2^+(q^++p_4^+)+p_3^+(p_1^+-q^+)]\over2p_1^+(p_1^+-q^+)(q^++p_4^+)
(p_{12}^+-q^+)^2}
\eea
Integrating over $q^-, {\bfs q}$ produces, assuming $p_{14}^+>0$, 
\bea
{\Gamma_{23}^{\wedge\wedge\wedge\vee}\over16g^4N_c^2}
&=&{K_{41}^{\wedge2}\over32\pi^2p_1^{+2}p_{14}^2}
\bigg\{\sum_{q^+<-p_4^+}{-q^+\over (p_1^+-q^+)}
\left[{p_2^+(p_4^++q^+)+p_3^+(p_1^+-q^+)
\over(q^++p_4^+)(q^+-p_{12}^+)^2}\right]
\nonumber\\
&&\hskip.8in + 
\sum_{p_1^+>q^+>-p_4^+}{p_4^{+}\over p_{14}^+}
\left[{p_2^+(p_4^++q^+)+p_3^+(p_1^+-q^+)
\over(q^++p_4^+)(q^+-p_{12}^+)^2}\right]
\bigg\}\nonumber\\
&=&{K_{41}^{\wedge2}\over32\pi^2p_1^{+2}p_{14}^2}
\bigg\{\sum_{q^+<-p_4^+}\left[-{p_1^+\over p_2^+(p_1^+-q^+)}
+{p_4^+
\over p_3^+(q^++p_4^+)}+{p_2^+p_4^++p_1^+p_3^+\over p_2^+p_3^+
(p_{12}^+-q^+)}+{2p_{12}^+\over(p_{12}^+-q^+)^2}\right]
\nonumber\\
&&\hskip.8in + 
\sum_{p_1^+>q^+>-p_4^+}
\left[
+{p_4^+\over p_3^+(q^++p_4^+)}+{p_4^+\over p_3^+
(p_{12}^+-q^+)}+{2p_{2}^+p_4^+\over p_{14}^+(p_{12}^+-q^+)^2}\right]
\bigg\}\nonumber\\
&\to&{K_{41}^{\wedge2}\over32\pi^2p_1^{+2}p_{14}^2}
\left[{p_1^+\over p_2^+}\ln{p_{12}^+p_{14}^+\over p_1^+|p_3^+|}
+{p_4^+\over p_3^+}\ln{p_{12}^+p_{14}^+\over p_2^+|p_4^+|}\right]
\eea
The coefficient of $16g^4N_c^2(16\pi^4q_0^2q_2^2q_3^2)^{-1}$ is
\bea
K_{35}^\wedge K_{64}^\wedge
{p_4^+[p_1^+(q^+-p_{12}^+)-p_2^+q^+)]\over2p_3^+(p_{12}^+-q^+)q^+
(p_{1}^+-q^+)^2}
\eea
and integration gives
\bea
{\Gamma_{12}^{\wedge\wedge\wedge\vee}\over16g^4N_c^2}
&=&{K_{34}^{\wedge2}\over32\pi^2p_3^{+2}p_{12}^+p_{12}^2}
\bigg\{\sum_{q^+<-p_4^+}{p_3^+\over (p_{12}^+-q^+)}
\left[{p_1^+(p_{12}^+-q^+)+p_2^+q^+
\over(q^+-p_1^+)^2}\right]
\nonumber\\
&&\hskip1.8in + \sum_{q^+>-p_4^+}{p_4^{+}\over q^+}
\left[{p_1^+(p_1^++p_2^+-q^+)+p_2^+q^+
\over(q^+-p_1^+)^2}\right]
\bigg\}\\
&=&{K_{34}^{\wedge2}\over32\pi^2p_3^{+2}p_{12}^+p_{12}^2}
\bigg\{\sum_{q^+<-p_4^+}\left[{p_3^+p_{12}^+\over p_2^+(p_{12}^+-q^+)}
+{2p_1^+p_3^+
\over(q^+-p_1^+)^2}+{p_3^+p_{12}^+\over p_2^+(q^+-p_{1}^+)}\right]
\nonumber\\
&&\hskip1.8in + \sum_{q^+>-p_4^+}\left[{p_{12}^+p_4^{+}\over p_1^+q^+}
+{2p_2^+p_4^+
\over(q^+-p_1^+)^2}+{p_{12}^+p_4^{+}\over p_1^+(p_1^+-q^+)}\right]
\bigg\}\\
&\to&{K_{34}^{\wedge2}\over32\pi^2p_3^{+2}p_{12}^2}
\bigg\{{1\over p_{12}^+}\sum_{q^+>-p_4^+}{2p_2^+p_4^+
\over(q^+-p_1^+)^2}-{2p_3^+p_4^+\over p_{12}^+p_{14}^+}
+{p_1^+p_3^++p_2^+p_4^+\over p_1^+p_2^+}
\ln{p_{12}^+p_{14}^+\over p_1^+p_2^+}
\bigg\}
\eea
The coefficient of $16g^4N_c^2(16\pi^4q_1^2q_2^2q_3^2)^{-1}$ is
\bea
&&K_{25}^\wedge K_{35}^\wedge\bigg\{
{(p_1^+-q^+)[p_4^+(q^+-p_{1}^+)-p_1^+(q^++p_4^+)]\over2p_2^+p_3^+
(q^++p_4^+)(p_{1}^++p_4^+)^2}\nonumber\\
&&\qquad-{q^++p_4^+\over p_1^+-q^+}\left[
{p_1^+(q^+-p_{1}^+)-p_4^+(q^++p_4^+)\over2p_2^+p_3^+(p_{1}^++p_4^+)^2}
+{p_1^+p_4^++(p_{1}^+-q^+)(q^++p_4^+)\over2p_2^+p_3^+q^{+2}}\right]\bigg\}
\eea
and integration gives
\bea
{\Gamma_{41}^{\wedge\wedge\wedge\vee}\over16g^4N_c^2}&=&
-{K_{23}^{\wedge2}\over 32\pi^2p_2^+p_3^+p_{12}^2}
\bigg\{\sum_{-p_4^+<q^+<p_1^+}{q^++p_4^+\over p_{14}^+}
\bigg[{q^++p_4^+\over p_1^+-q^+}\nonumber\\&&
\hskip-.7in\left({(q^+-p_1^+)p_1^+-(q^++p_4^+)p_4^+\over(p_1^++p_4^+)^2}
+{(q^++p_4^+)(p_1^+-q^+)+p_1^+p_4^+
\over q^{+2}}\right)
-{p_1^+-q^+\over q^++p_4^+}{(q^+-p_1^+)p_4^+-(q^++p_4^+)p_1^+
\over(p_1^++p_4^+)^2}\bigg]\nonumber\\
&&\hskip-.7in+\sum_{q^+>p_1^+}{p_{12}^+-q^+\over p_2^+}
\bigg[{q^++p_4^+\over p_1^+-q^+}
\left({(q^+-p_1^+)p_1^+-(q^++p_4^+)p_4^+\over(p_1^++p_4^+)^2}
+{(q^++p_4^+)(p_1^+-q^+)+p_1^+p_4^+
\over q^{+2}}\right)\nonumber\\
&&\hskip.5in-{p_1^+-q^+\over q^++p_4^+}{(q^+-p_1^+)p_4^+-(q^++p_4^+)p_1^+
\over(p_1^++p_4^+)^2}\bigg]\bigg\}
\eea
The coefficient of $16g^4N_c^2(16\pi^4q_0^2q_1^2q_2^2)^{-1}$ is
\bea
&&K_{61}^\wedge K_{25}^\wedge\bigg\{
{(p_{12}^+-q^+)[p_4^+(p_{12}^+-q^+)+p_3^+q^+]\over2p_1^+p_2^+
q^+(p_{1}^++p_2^+)^2}\nonumber\\
&&\qquad-{q^+\over p_{12}^+-q^+}\left[
{q^+(p_{12}^+-q^+)+p_3^+p_4^+\over2p_1^+p_2^+(q^++p_4^+)^2}
+{q^+p_4^++(p_{12}^+-q^+)p_3^+\over2p_1^+p_2^+(p_1^++p_2^+)^{ 2}}\right]\bigg\}
\eea
and integration gives
\bea
{\Gamma_{34}^{\wedge\wedge\wedge\vee}\over16g^4N_c^2}&=&
-{K_{12}^{\wedge2}\over 32\pi^2p_{1}^+p_2^+p_{12}^+
p_{12}^2}
\bigg\{
\sum_{q^+<p_1^+}\bigg[{q^{+2}\over p_1^+(p_{12}^+-q^+)}
\left({q^+p_4^++(p_{12}^+-q^+)p_3^+\over(p_1^++p_2^+)^2}
+{q^+(p_{12}^+-q^+)+p_3^+p_4^+
\over(q^++p_4^+)^2}\right)\nonumber\\
&&-{(p_{12}^+-q^+)\over p_1^+}
{(q^+p_3^++(p_{12}^+-q^+)p_4^+)
\over(p_1^++p_2^+)^2}\bigg]\nonumber\\
&&+\sum_{q^+>p_1^+}\bigg[{q^+\over p_2^+}
\left({q^+p_4^++(p_{12}^+-q^+)p_3^+\over(p_1^++p_2^+)^2}
+{q^+(p_{12}^+-q^+)+p_3^+p_4^+
\over(q^++p_4^+)^2}\right)\nonumber\\&&
-{(p_{12}^+-q^+)^2\over p_2^+q^+}
{(q^+p_3^++(p_{12}^+-q^+)p_4^+)
\over(p_1^++p_2^+)^2}\bigg]\bigg\}
\eea

\subsection{Result and discussion}
 Since we already know 
the results of cubic vertex corrections 
and self-energy insertion on  internal lines, let us concentrate 
here on 
the triangle-like contributions originating from the box reduction.
Due to kinematic constraints on $q^+$, 
different diagrams live in different patches on the worldsheet.
For the discussion let us assume without loss of generality that 
$p_1^+,~p_2^+>0$ 
and $p_1^++p_4^+>0$. The arguments go similarly 
for the case $p_1^++p_4^+<0$. Then we divide 
the whole worldsheet into three patches $0\le q^+\le -p_4^+$, 
 $-p_4^+\le q^+\le p_1^+$ and  $p_1^+\le q^+\le p_{12}^+$.  
Diagrams with propagators
 $(q_0^2q_1^2q_3^2)^{-1}$ can live only in first two regions ($q^+<p_1^+$) 
while the diagrams 
with $(q_1^2q_2^2q_3^2)^{-1}$ live in the second and 
third regions ($q^+>-p_4^+$). The diagrams with 
 $(q_0^2q_2^2q_3^2)^{-1}$ and  $(q_0^2q_1^2q_2^2)^{-1}$ 
live in all three regions.  For  the calculation, 
we use Schwinger 
parameters for the internal propagators and 
UV cutoff $\delta$ as discussed in appendix \ref{boxintegral} 
and use  {\it Mathematica} to 
carry out the algebra.
To see the cancellation of divergences, 
it is very helpful to convert all the polarization structures 
into two independent forms by using the $K$-identities 
before summing over $q^+$. Using the $K$-identities, 
we  first express all the $K_{ij}^\wedge$ and $K_{ij}^\vee$ 
in terms of $K_{12}^\wedge$ and $K_{34}^\wedge$.
 With $K_{12}^\wedge$ and
 $K_{34}^{\wedge }$, we can form three bilinear 
structures $K_{12}^{\wedge 2}$, $K_{34}^{\wedge 2}$ and 
$K_{12}^\wedge K_{34}^\wedge$. But, these three bilinears 
are again related to each other by
\bea
{K_{12}^\wedge K_{34}^\wedge \over p_{12}^2}=-{K_{32}^\wedge K_{41}^\wedge 
\over p_{14}^2}.\label{strelation}
\eea
Using 
\bea
K_{32}^\wedge&=&{p_3^+ K_{12}^\wedge-p_2^+ K_{34}^\wedge \over p_{12}^+}, \\
K_{41}^\wedge&=& -{p_4^+ K_{12}^\wedge-p_1^+ K_{34}^\wedge \over p_{12}^+}
\eea 
in Eq. (\ref{strelation}) we can eliminate $K_{34}^{\wedge 2}$  by
\bea
K_{34}^{\wedge 2}=-{p_3^+p_4^+ \over p_1^+p_2^+}K_{12}^{\wedge 2}
+{(p_1^+p_3^++p_2^+p_4^+)p_{12}^2
+p_{12}^{+2} p_{14}^2 \over p_1^+p_2^+ p_{12}^2}K_{12}^\wedge K_{34}^\wedge.
\eea
So,  we have only two independent polarizations $K_{12}^{\wedge 2}$ 
and $K_{12}^\wedge K_{34}^\wedge$.

 After integrating over $q^-,~{\bf q}$ and the parameter $T=T_1+T_2+T_3$,
 for each triangle-like term  we have 
an expression of the form
\bea
&&\sum_{q^+}\int dx_1 dx_2 dx_3\delta(1-x_1-x_2-x_3)\delta(f(q^+,x_i,p_i^+))
\big\{K_{12}^{\wedge 2} ( B_1+ B_1^\prime 
\ln(\delta e^\gamma H(x_i x_j p^2)) )+ \nonumber \\
&& +K_{12}^{\wedge}K_{34}^{\wedge}(B_2+ B_2^\prime 
\ln(\delta e^\gamma H(x_i x_j p^2)))+ 
K_{12}^{\wedge} k_i^\wedge B_3^i+K_{34}^{\wedge} k_i^\wedge B_4^i \big\},
\eea
where $x_i={T_i/T}$.  $k_i$ are the 
dual momenta assigned to the external regions 
of the loop as shown in Fig. \ref{boxreduction1} and $B_i$
and $B_i^\prime$ are functions of $q^+, p_i^+$ and $x_i$.
 One can next peacefully perform
 all the integrations except that over $q^+$ which we 
discretize to take care of the singularities. 
Individually, each diagram is divergent in $q^+$ 
in the UV cutoff $\delta$, and does not look simple. But amazing 
simplifications occur when all diagrams are combined. 
The physical scattering amplitude $\Gamma^{\wedge\wedge\wedge\vee}$ 
should be proportional to two powers 
of $K_{ij}^\wedge$. But in each triangle-like diagram,  $B_3$ and  $B_4$ 
are proportional to only one power of $K_{ij}^\wedge$ 
and one power of $k_i^\wedge$ and are not of that form. 
Nevertheless, when all 
the allowed triangle-like diagrams in each region on the worldsheet 
are put together, 
 after integration over the $x_i$'s, the $B_3$ and  $B_4$ terms 
combine nicely  to have the appropriate bilinear structures 
in the $K_{ij}^\wedge$'s. 

From the expressions of the triangle-like 
diagrams (Eqs. (\ref{T013})-(\ref{T023})), we can see that the
four possible places where we can encounter 
divergences are  $q^+=0,q^++p_4^+=0,q^+-p_1^+=0$ 
and $p_{12}^+-q^+=0$. The apparent divergences at 
the end points $q^+=0$ and $p_{12}^+-q^+=0$ are 
tamed by the limits of integration over the Feynamn parameters $x_i$, 
and we need only worry about the singularities at the interior 
points $q^++p_4^+=0,q^+-p_1^+=0$. 
 In the first patch of the worldsheet, 
Eq. (\ref{T013}) and Eq. (\ref{T023})  
individually are linearly divergent at $q^++p_4^+=0$. One
 part of this divergent term in Eq. (\ref{T013}) 
cancels the divergence of  Eq. (\ref{T023}) and 
the other part
is canceled by the similar term in the quartic 
triangle $\Gamma_{34}^{\wedge\wedge\wedge\vee}$.
Similarly in all the three regions,   
when triangle-like diagrams are combined 
with  the quartic triangle diagrams,
all the linear divergences  at $q^+-p_1^+=0$ and $q^++p_4^+=0$ 
cancel out and we can take the continuum limit of the $q^+$ sums 
and perform the resulting integration over $q^+$. 
We also take out the logarithm terms with polynomial coefficients
and integrate over $q^+$. But, the logarithm terms 
whose  coefficients have $q^+$ in the denominators 
are singular and we keep sum over discrete $q^+$ for them.
 Those terms are canceled when
 added with $\Gamma_\Delta^{\wedge\wedge\wedge\vee}$ and 
  $\Gamma_{SE}^{\wedge\wedge\wedge\vee}$.    
All the triangle-like  and  quartic triangle terms combine nicely to produce 
\bea
\Gamma_{TL}^{\wedge\wedge\wedge\vee}
={(2g)^4N_c^2\over 32\p^2}\bigg\{
B_0 +{p_4^+\over p_1^+p_2^+p_3^+} 
{K_{12}^\wedge K_{34}^\wedge \over p_{12}^2} \bigg [
{11\over 3} \ln(\delta e^\gamma p_{12}^2)-{11\over 3} 
\ln(\delta e^\gamma p_{14}^2)-S^{q^+}(p_i^+) \bigg ]\bigg \},
\eea
where $B_0$ is given by
\bea
B_0&=&\bigg [-{p_1^+(p_2^+-p_3^+)-3p_3^+(p_2^++p_3^+) 
\over 3 p_2^+p_{12}^{+}p_{23}^+p_4^+ p_{14}^2}
-{p_3^+(-3p_1^+p_{12}^++p_3^+(-p_1^++p_2^+)) 
\over 3 p_1^+p_2^+p_{12}^{+2}p_4^+ p_{12}^2 } \bigg ] 
K_{12}^{\wedge 2}\nonumber \\
&~&+\bigg [ { p_1^+(-p_2^++p_3^+)+p_3^+p_{23}^+ 
\over 3p_1^+p_3^+p_{23}^{+2} p_{12}^2}-
{p_1^+(-p_2^++p_3^+)+3p_3^+p_{23}^+ 
\over 3 p_{12}^+p_3^+p_{23}^+p_4^+ p_{14}^2}\bigg ]K_{12}^\wedge K_{34}^\wedge
\eea
and $S^{q^+}(p_i^+)$ contains the terms with logarithms. 
With a bit rearrangement,  it can be written as 
\bea
S^{q^+}(p_i^+)&=&S_3^{q^+}(p_1,p_2)
+S_3^{q^+}(-p_4,-p_3)-S_2^{q^+}(-p_4,-p_{23})\nonumber \\
&~&-S_1^{q^++p_4^+}(p_{14},p_2)
+2 {\cal A}^{q^+}(p_{12}^2,p_{12}^+)
-2{\cal A}^{q^++p_4^+}(p_{14}^2,p_{14}^+).
\eea
The physical scattering amplitude 
is obtained by adding the box and quartic triangles with all 
the cubic vertex corrections 
and the self energy insertions on the internal lines. 
Then we are just left with
\bea
\Gamma^{\wedge\wedge\wedge\vee}&=&\Gamma_{TL}^{\wedge\wedge\wedge\vee}
+\Gamma_{\Delta}^{\wedge\wedge\wedge\vee}
+\Gamma_{SE}^{\wedge\wedge\wedge\vee}\nonumber \\
&=&(2g)^4N_c^2\bigg\{ {B_0\over 32 \p^2}
 -{p_4^+\over32\pi^2p_1^+p_2^+p_3^+}
{K_{34}^\wedge K_{12}^\wedge\over p_{12}^2}\bigg 
[{p_1^+p_2^+\over 3p_{12}^{+2}}-
{p_2^+p_3^+\over 3p_{14}^{+2}}\bigg ]
\nonumber\\
&~&+{p_3^+\over48\pi^2p_1^+p_2^+p_4^+p_{12}^{+2}}
{K_{12}^{\wedge3}K_{34}^\vee\over p_{12}^4}
+{p_1^+\over48\pi^2p_2^+p_3^+p_4^+p_{14}^{+2}}
{K_{23}^{\wedge3}K_{41}^\vee\over p_{14}^4}\bigg\}.
\eea
The above expression  simplifies to
\bea
\Gamma^{\wedge\wedge\wedge\vee}&=&{(2g)^4N_c^2\over 32 \p^2}
\bigg [{p_2^+ p_3^+ p_{14}^4+p_1^+ 
(p_2^+ p_{12}^4+p_3^+(p_{12}^2+p_{14}^2)^2) \over 
 3 p_1^+p_2^+ p_{12}^+p_4^+ p_{12}^4 p_{14}^4} K_{12}^{\wedge 2}\nonumber \\
&~& -
{(p_2^+ +p_3^+ )p_{12}^2+(p_1^++p_2^++2 p_3^+)p_{14}^2 \over
 3 p_3^+ p_{12}^+p_4^+ p_{12}^2 p_{14}^4} K_{12}^\wedge K_{34}^\wedge 
\bigg](p_{12}^2+p_{14}^2)
\eea
which again can be rewritten in the compact form
\bea
\Gamma^{\wedge\wedge\wedge\vee}=
-{g_s^4N_c^2\over 96 \p^2}{p_2^+ p_4^+ K_{13}^{\wedge 2} 
\over K_{43}^\wedge K_{32}^\vee 
K_{21}^\vee K_{14}^\wedge}(p_{12}^2+p_{14}^2)
\eea
agreeing with the known result \cite{bernk,kunsztst} after removing a factor
of $N_c$.
\vskip14pt
\noindent\underline{ Acknowledgments}: 
We would like to thank Zvi Bern for sharing his valuable insight
into the structure of gauge theory loop integrands. This work was
initiated during the KITP program on QCD and String Theory in
Fall 2004, and CBT thanks the Kavli Institute and all participants for 
providing a stimulating environment. 
We also thank Lisa Everett and Charlie Sommerfield
for critical comments on the manuscript.
This research was supported in part by the Department
of Energy under Grant No. DE-FG02-97ER-41029.

\appendix
\section{Evaluation of Loop Momentum Integrals}\label{boxintegral}
The loop integrals we encounter 
in this article can be most easily handled through the
introduction of Schwinger parameters $T_1,T_2,T_3,T_4$ for the
internal line propagators 
$(q-k_0)^{-2}, (q-k_1)^{-2},(q-k_2)^{-2},(q-k_3)^{-2}$ respectively.
For the helicity nonconserving processes we have shown that
box integrals can always be reduced to triangle-like integrals,
which means we will only be integrating three of these parameters
setting the fourth to zero. However keeping all four $T$'s
allows us to handle in a unified way the four different
triangle topologies we require. We only need remember to set
the appropriate one to zero when we do each triangle-like integral.
Since some of the diagrams are divergent in the ultraviolet, we also retain
the worldsheet UV cutoff factors $e^{-\delta{\bfs q}^2}$. The integral
over $q^-$ is equivalent to the
insertion of a delta function
\bea
\int dq^-\to \pi\delta(T_{14}q^+-T_2p_1^+-T_3p_{12}^+
+T_4p_4^+)
\eea
where we use the shorthand $T_{14}=T_1+T_2+T_3+T_4$.
The integration over
${\bfs q}$ is then a Gaussian that is easily done by completing the
square and shifting ${\bfs q}\to{\bfs q}+{\bfs K}$, with
\bea
{\bfs K}={{\bfs k}_0 T_1 +{\bfs k}_1 T_2 +{\bfs k}_2 T_3 +{\bfs k}_3 T_4 \over
T_{14}+\delta}
\eea 
 One then finds, 
using the Feynman parameters $x_i\equiv T_i/T_{14}$ that
\bea
{\bfs K}_{16}&\to& -p_1^+{\bfs q}+q^+{\bfs p}_1
-x_3{\bfs K}_{12}-x_4{\bfs K}_{41}
+p_1^+{\delta{\bfs K}\over T_{14}}\\
{\bfs K}_{52}&\to& -p_2^+{\bfs q}+q^+{\bfs p}_2
-x_4{\bfs K}_{23}-x_1{\bfs K}_{12}
+p_2^+{\delta{\bfs K}\over T_{14}}\\
{\bfs K}_{35}&\to& p_3^+{\bfs q}-q^+{\bfs p}_3
+x_2{\bfs K}_{23}+x_1{\bfs K}_{34}
-p_3^+{\delta{\bfs K}\over T_{14}}\\
{\bfs K}_{64}&\to& p_4^+{\bfs q}-q^+{\bfs p}_4
+x_3{\bfs K}_{34}+x_2{\bfs K}_{41}
-p_4^+{\delta{\bfs K}\over T_{14}}
\eea
We shall have use for the following combinations of momenta which
arise in the loop integrand after shifting $q$ and
sending $\delta\to0$ 
\bea
K_0&=&x_2p_1+x_3(p_1+p_2)-x_4p_4\\
K_0-p_1&=&x_3p_2+x_4(p_2+p_3)-x_1p_1\\
K_0-p_1-p_2&=&x_4p_3+x_1(p_3+p_4)-x_2p_2\\
K_0+p_4&=&x_1p_4+x_2(p_1+p_4)-x_3p_3
\eea
It is convenient to change variables to $T=T_{14}$ and the $x_i$ after
which the $T$ integral can be done. 
In the evaluation of triangle-like diagrams there are two $x_i$ to 
integrate. The $q^+$ integral is 
discretized and the corresponding sum over $q^+$ is always
done last.
In the evaluation of the on-shell  triangle diagram, we encounter integrals 
of the form
\bea
\int_{x+y\leq1} dx dy\delta(q^+-(x+y)p^+_1-yp^+_2){\cal I}
\eea
where the integrand is a linear function of $x$ times a linear function of
$\ln xy$, $\ln x(1-x-y)$, or $\ln y(1-x-y)$. By $p^+$ conservation,
two of the momenta $p^+_{1,2,3}$ have one sign and the 
third has the opposite sign. In this section we label momenta so that
$p_1^+>0$ and $p_3^+<0$ have the same sign. If $p_2^+$
is positive do the above integral in its displayed form. If $p_2^+$
is negative rewrite the argument of the delta function in terms
of $p_2^+=-|p_2^+|$ and $p_3^+=-|p_3^+|$, and 
rename $x\leftrightarrow y$, which brings the integral to the form
\bea
\int_{x+y\leq1} dx dy
\delta(q^+-(x+y)|p^+_3|-y|p^+_2|){{\cal I}}
\eea
which reduces it to the first form, with $|p^+_3|$ in the
role of $p_1^+$ and $|p_2^+|$ in the role of $p_2^+$.  
Thus, without loss of generality we
can stipulate that $p_1^+,p_2^+>0$.
Then we do the $y$ integral which sets 
$y=(q^+-xp_1^+)/p^+_{12}$, and also sets
the range of the $x$ integral $0<x<x_m$ where $x_m=q^+/p_1^+$ for
$q^+<p_1^+$ and $x_m=(p^+_{12}-q^+)/p_2^+$ for $q^+>p_1^+$.
Then the following $x$ integrals are needed:
\bea
\hskip-.3in \int dx\ \ln(xy)&=&\left(x_m-{q^+\over p_1^+}\right)
\ln{q^+-x_mp_1^+\over p^+_{12}}
+{q^+\over p_1^+}\ln{q^+\over p^+_{12}}+x_m\ln x_m-2x_m\noindent\\
&=&\cases{\displaystyle
{q^+\over p_1^+}\left(\ln{q^+\over p^+_{12}}+\ln{q^+\over p_1^+}
-2\right)
& for $q^+<p_1^+$\cr
\phantom{[}&\cr
\displaystyle
{p^+_{12}(p_1^+-q^+)\over p_1^+p_2^+}
\ln{q^+-p_1^+\over p_2^+}
+{q^+\over p_1^+}\ln{q^+\over p^+_{12}}&\cr
\displaystyle\hskip1in +{p^+_{12}-q^+\over p_2^+}
\left(\ln{p^+_{12}-q^+\over p_2^+} -2\right)
& for $q^+>p_1^+$\cr}\\
\hskip-.3in \int dx\ x\ln(xy)
&=&{1\over2}\left(x_m^2-{q^{+2}\over p_1^{+2}}\right)
\ln{q^+-x_mp_1^+\over p^+_{12}}
+{q^{+2}\over 2p_1^{+2}}\ln{q^+\over p^+_{12}}+{x_m^2\over2}\ln x_m-
{x_m^2p_1^++q^+x_m\over2p_1^+}\nonumber\\
&=&\cases{\displaystyle
{q^{+2}\over 2p_1^{+2}}\left(\ln{q^+\over p^+_{12}}+\ln{q^+\over p_1^+}
-2\right)
& for $q^+<p_1^+$\cr
\phantom{[}&\cr
\displaystyle
{p^+_{12}(p_1^+-q^+)\over 2p_1^+p_2^+}\left[
{q^+\over p_1^+}+{p^+_{12}-q^+\over p_2^+}\right]
\ln{q^+-p_1^+\over p_2^+}
+{q^{+2}\over 2p_1^{+2}}\ln{q^+\over p^+_{12}}\hskip-.5in&\cr
\displaystyle\quad +{(p^+_{12}-q^+)^2\over 2p_2^{+2}}
\left(\ln{p^+_{12}-q^+\over p_2^+} -1\right)
-{q^+(p^+_{12}-q^+)\over 2p_1^+p_2^{+}}
& for $q^+>p_1^+$\cr}
\eea
\bea
\hskip-.3in \int dx\ \ln x(1-x-y)
&=&\cases{\displaystyle
{q^+\over p_1^+}\left(\ln{q^+\over p_1^+}-2\right)
+{p^+_{12}-q^+\over p_2^+}
\ln{p^+_{12}-q^+\over p^+_{12}}&\cr
\displaystyle\hskip1in -\ 
{p^+_{12}(p_1^+-q^+)\over p_1^+p_2^+}
\ln{p_1^+-q^+\over p_1^+}
& for $q^+<p_1^+$\cr
\phantom{[}&\cr
\displaystyle
{p^+_{12}-q^+\over p_2^+}
\left(\ln{p^+_{12}-q^+\over p^+_{12}}
+\ln{p^+_{12}-q^+\over p_2^+} -2\right)
& for $q^+>p_1^+$\cr}\\
&&\phantom{A}\nonumber\\
\hskip-.3in \int dx\ x\ln(x(1-x-y))&=&
\cases{\displaystyle
{q^{+2}\over 2p_1^{+2}}\left(\ln{q^+\over p_1^+}-1\right)
+{(p^+_{12}-q^+)^2\over 2p_2^{+2}}
\ln{p^+_{12}-q^+\over p^+_{12}}
-{q^+(p^+_{12}-q^+)\over 2p_1^+p_2^{+}}
\hskip-.5in&\cr
\displaystyle\quad 
-{p^+_{12}(p_1^+-q^+)\over 2p_1^+p_2^+}\left[
{q^+\over p_1^+}+{p^+_{12}-q^+\over p_2^+}\right]
\ln{p_1^+-q^+\over p_1^+}
& for $q^+<p_1^+$\cr
\phantom{[}&\cr
\displaystyle
{(p^+_{12}-q^+)^2\over 2p_2^{+2}}
\left(\ln{p^+_{12}-q^+\over p^+_{12}}+
\ln{p^+_{12}-q^+\over p_2^+} -2\right)
& for $q^+>p_1^+$\cr}
\eea
\bea
\hskip-.3in\int dx\ \ln y(1-x-y)&=&\cases{\displaystyle
{q^+\over p_1^+}\left(\ln{q^+\over p^+_{12}}-2\right)
+{p^+_{12}-q^+\over p_2^+}
\ln{p^+_{12}-q^+\over p^+_{12}}&\cr
\displaystyle\hskip1in -\ 
{p^+_{12}(p_1^+-q^+)\over p_1^+p_2^+}
\ln{p_1^+-q^+\over p_1^+}
& for $q^+<p_1^+$\cr
\phantom{[}&\cr
\displaystyle
{q^+\over p_1^+}\ln{q^+\over p^+_{12}}+{p^+_{12}-q^+\over p_2^+}
\left(\ln{p^+_{12}-q^+\over p^+_{12}} -2\right)
&\cr
\displaystyle\hskip1in -\ 
{p^+_{12}(q^+-p_1^+)\over p_1^+p_2^+}
\ln{q^+-p_1^+\over p_2^+}
& for $q^+>p_1^+$\cr}\\
\hskip-.3in \int dx\ x\ln(y(1-x-y))&=&
\cases{\displaystyle
{q^{+2}\over 2p_1^{+2}}\left(\ln{q^+\over p^+_{12}}-2\right)
+{(p^+_{12}-q^+)^2\over 2p_2^{+2}}
\ln{p^+_{12}-q^+\over p^+_{12}}
-{q^+(p^+_{12}-q^+)\over 2p_1^+p_2^{+}}
\hskip-.5in&\cr
\displaystyle\quad 
-{p^+_{12}(p_1^+-q^+)\over 2p_1^+p_2^+}\left[
{q^+\over p_1^+}+{p^+_{12}-q^+\over p_2^+}\right]
\ln{p_1^+-q^+\over p_1^+}
& for $q^+<p_1^+$\cr
\phantom{[}&\cr
\displaystyle
{q^{+2}\over 2p_1^{+2}}\ln{q^+\over p^+_{12}}
+{(p^+_{12}-q^+)^2\over 2p_2^{+2}}
\left(\ln{p^+_{12}-q^+\over p^+_{12}} -2\right)
-{q^+(p^+_{12}-q^+)\over 2p_1^+p_2^{+}}
\hskip-.5in&\cr
\displaystyle\quad 
-{p^+_{12}(q^+-p_1^+)\over 2p_1^+p_2^+}\left[
{q^+\over p_1^+}+{p^+_{12}-q^+\over p_2^+}\right]
\ln{q^+-p_1^+\over p_2^+}& for $q^+>p_1^+$ \cr}
\eea

\end{document}